\documentclass[fleqn,10pt]{wlscirep}
\usepackage[utf8]{inputenc}
\usepackage[T1]{fontenc}
\usepackage{color, colortbl}
\definecolor{Gray}{gray}{0.9}
\title{Critical behavior and magnetocaloric effect	across the magnetic transition in Mn$_{1+x}$Fe$_{4-x}$Si$_{3}$}

\author[1,2*]{Vikram Singh}
\author[1]{Pallab Bag}
\author[2]{R. Rawat}
\author[1**]{R. Nath}
\affil[1]{School of Physics, Indian Institute of Science Education and Research, Thiruvananthapuram-695551, India}
\affil[2]{UGC-DAE Consortium for Scientific Research, University Campus, Khandwa Road, Indore-452001, India}

\affil[*]{vikram51128@gmail.com}

\affil[**]{rnath@iisertvm.ac.in}

\begin{abstract}
  The nature of the magnetic transition, critical scaling of magnetization, and magnetocaloric effect in Mn$_{1+x}$Fe$_{4-x}$Si$_{3}$ ($ x =$ 0 to 1) are studied in detail. Our measurements show no thermal hysteresis across the magnetic transition for the parent compound which is in contrast with the previous report and corroborate the second order nature of the transition. The magnetic transition could be tuned continuously from 328~K to 212~K with Mn substitution at the Fe site. The Mn substitution leads to a linear increase in the unit cell volume and a slight reduction in the effective moment. A detailed critical analysis of the magnetization data for $x = 0.0$ and 0.2 is performed in the critical regime using the modified Arrott plots, Kouvel-Fisher plot, universal curve scaling, and scaling analysis of magnetocaloric effect. The magnetization isotherms follow modified Arrott plots with critical exponent ($\beta \simeq 0.308$, $\gamma \simeq 1.448$, and $\delta \simeq 5.64$) for the parent compound ($x=0.0$) and ($\beta \simeq 0.304$, $\gamma \simeq 1.445$, and $\delta \simeq 5.64$) for $x = 0.2$. The Kouvel-Fisher and universal scaling plots of the magnetization isotherms further confirm the reliability of our critical analysis and values of the exponents. These values of the critical exponents are found to be same for both the parent and doped samples which do not fall under any of the standard universality classes. The exchange interaction decays as $J(r)\sim r^{-3.41}$ following the renormalization group theory and the observed critical exponents correspond to lattice dimensionality $d=2$, spin dimensionality $n=1$, and the range of interaction $\sigma=1.41$. This value of $\sigma (<2 )$ indicates long-range interaction between magnetic spins. A reasonable magnetocaloric effect $\Delta S_{\rm m}\simeq-6.67$~J/Kg-K and -5.84~J/Kg-K for $x= 0.0$ and 0.2 compounds, respectively, with a huge relative cooling power ($RCP \sim 700$~J/Kg) for 9~T magnetic field change is observed. The universal scaling of magnetocaloric effect further mimics the second order character of the magnetic transition. The obtained critical exponents from the critical analysis of magnetocaloric effect agree with the values deduced from the magnetic isotherm analysis.
\end{abstract}
\begin{document}

\flushbottom
\maketitle
\thispagestyle{empty}
\section*{Introduction}
The research on magnetic materials with large magnetocaloric effect (MCE) has increased immensely in recent past since such materials could be used for magnetic refrigeration, an alternative to conventional vapor compression technique.\cite{Franco112,GschneidnerJr1479,Tishin2016} The MCE is defined as the isothermal change in magnetic entropy or adiabatic change in temperature with change in external magnetic field, which generally has large value across the magnetic phase transitions. The nature of the magnetic phase transition essentially plays an important role in deciding the practical use of the materials. The giant MCE is observed in various materials across the first order magnetic phase transition due to strong coupling between electronic, structural, and magnetic degrees of freedom.\cite{Pecharsky44,Pecharsky4494,Gschneidner387,Moya439,Roy183201,Tegus150} However, the drawback of first order phase transition in comparison to second order transition is the hysteresis losses. Therefore, second order phase transition with large MCE could be favorable for magnetic refrigeration purpose where system has to go through repeated cycling.\cite{Chaikin1995,Stanley336,Li152403,Singh3321} 
Further, for the application purpose, materials with large MCE near room temperature are desirable and rare earth based intermetallic systems due to their large magnetic moment are prominent in the list. However, the high cost of rare earth elements often restricts the use of these materials.\cite{Pecharsky4494,GschneidnerJr1479,Gschneidner945} Therefore, the transition metal based intermetallic compounds with large magnetic moment are widely preferred for this purpose.\cite{Bruck763,Gschneidner945,Gutfleisch821,Chirkova15,Tegus150,Krenke450,Singh3321}

In this regard, MnFe$_{4}$Si$_{3}$, which belongs to the Mn$_{5-x}$Fe$_{x}$Si$_{3}$ ($x = 0$ to 5) family, is a potential candidate because of its near room temperature paramagnetic (PM)-ferromagnetic (FM) transition accompanied with a large change in magnetization. The series Mn$_{5-x}$Fe$_{x}$Si$_{3}$ ($x = 0$ to 5) exhibits multiple magnetic phase transitions over a wide temperature range and MCE is observed across these transitions.\cite{Gottschilch15275,Gourdon56,Songlin249,Johnson311,Candini6819} 
In this series, one end compound Mn$_{5}$Si$_{3}$ undergoes two successive magneto-structural transitions: one from paramagnetic (PM) to collinear antiferromagnetic (AF2) state at $T_{\rm N2} \sim 100$~K coupled with a hexagonal to orthorhombic distortion followed by a AF2 to non-collinear antiferromagnetic (AF1) state at a lower temperature $T_{\rm N1} \sim 65$~K coupled with an orthorhombic to monoclinic structural change. This system has been studied extensively due to its complex phase diagram, large topological hall resistance, and spin fluctuation driven large MCE across the field induced transitions at low temperature.\cite{Brown7619,Gottschilch15275,Surgers055604,Surgers3400,Biniskos257205} The Fe substitution at the Mn site shifts $T_{\rm N2}$ weakly towards high temperatures while $T_{\rm N1}$ remains almost unchanged for $x \leq 3.5$. However, for larger doping concentrations ($x > 3.5$), the transition at $T_{\rm N1}$ collapses and the AFM transition at $T_{\rm N2}$ is transformed to a FM one.\cite{Songlin249,Candini6819}
On the other hand, the compound at the other end of this series i.e. Fe$_{5}$Si$_{3}$, shows only a PM to FM transition above room temperature ($T_{\rm C}\simeq 370$~K). Unfortunately, Fe$_{5}$Si$_{3}$ is unstable below $800~^{o}$C and decomposes into Fe$_{3}$Si and FeSi within few hours time.\cite{Johnson311,Shinjo797}

MnFe$_{4}$Si$_{3}$ crystallizes in a hexagonal crystal structure with space group $P6_{3}/mcm$ at room temperature. Transition metal atoms occupy two different crystallographic sites $M1$ and $M2$ with Wyckoff positions 4d and 6g, respectively.\cite{Johnson465,Narasimhan1511,Johnson311,Binczycka_K13} The $M1$ site is fully occupied by the Fe atom, whereas the $M2$ site is shared by Fe (2/3) and Mn (1/3) atoms.
Recent neutron and x-ray diffraction studies on single crystals reveal that MnFe$_{4}$Si$_{3}$ crystallizes with a lower symmetry of $P\bar{6}$ where the transition metal atoms can have four inequivalent sites: $M1a$, $M1b$, $M2a$, and $M2b$.\cite{Hering7128} The $M1$ site is partially occupied by both Fe and Mn atoms while the $M2$ site is fully occupied by the Fe atoms. Nevertheless, $P6_{3}/mcm$ still can be considered as an average structure of the low symmetry space group $P\bar{6}$ with an assumption that $M1$ and $M2$ split into two sites [($M1a$, $M1b$) and ($M2a$ and $M2b$)] each. The magnetic structure refinement confirms that only the $M1$ site possesses the magnetic moment ($\sim 1.5~\mu_{\rm B}$/metal atom) and is ordered in the $ab$-plane.\cite{Hering7128} These observations are in contrast with the previous studies where all the transition metals are considered to have magnetic moment aligned along the $c$-axis.\cite{Binczycka_K13,Johnson311,Johnson465,Haug539,Narasimhan1511,Gourdon56} Interestingly, Hiring~et~al\cite{Hering7128} observed an anisotropic variation of lattice parameters with temperature without any change in crystal symmetry and a thermal hysteresis across the magnetic transition. On these bases, the phase transition was characterized as a first order type. In the subsequent studies using M\"{a}ssbauer spectroscopy and MCE, Herlitschke~et~al\cite{Herlitschke094304} found that the magnetic transition cannot be strictly characterized either as first order or second order type. Therefore, they proposed that this uncertainty could be due to the presence of Landau tricritical point near the magnetic transition.

Thus, the ambiguity about the nature of the transition in MnFe$_{4}$Si$_{3}$ and the possibility to tune the transition upon Mn substitution at the Fe site persuade us to re-examine the Mn$_{1+x}$Fe$_{4-x}$Si$_{3}$ series. We show that the PM to FM transition is second order in nature, in contrast to previous reports.\cite{Hering7128,Herlitschke094304,Biniskos104407} A detailed investigation of the PM to FM transition has been performed for $x = 0.0$ and 0.2 via critical analysis of the magnetization data and MCE studies to understand the nature of magnetic interaction.

\section*{Methods}
A series of polycrystalline Mn$_{1+x}$Fe$_{4-x}$Si$_{3}$ (with  $ x = 0.0, 0.2, 0.6, 0.4, 0.8 $, and $ 1.0 $) samples is synthesized by arc melting of the constituent elements of purity better than $99.98$~\% in a water cooled copper hearth in Ar atmosphere. The ingots thus obtained are flipped and re-melted four times to ensure the homogeneous mixing of elements. The weight loss after melting is estimated to be less than $\sim 1$\% of the total sample weight. Obtained ingots are wrapped in Ta foils for thermal annealing in vacuum at 950~$^{o}$C for five days followed by quenching in ice cooled water. The initial characterization to check the phase purity of all the samples is carried out by powder x-ray diffraction (XRD) with Cu K$_{\alpha}$ lab source ($\lambda = 1.5406$~\AA, PANalytical X’Pert Pro diffractometer). Temperature ($T$) dependent powder XRD measurements are carried out over a temperature range 300~K to 15~K for the Mn$_{2}$Fe$_{3}$Si$_{3}$ sample. For this purpose, an Oxford PheniX closed-cycle helium cryostat is used as an attachment to the diffractometer. The synchrotron powder XRD (SXRD) measurement for the parent MnFe$_{4}$Si$_{3}$ sample is performed to detect the presence of minor secondary phase of FeMn as reported previously.\cite{Hering7128} It is carried out at the angle dispersive x-ray diffraction (ADXRD) beamline (BL-12), Indus-2 synchrotron source, RRCAT.\cite{Sinha072017}
The calibration of photon energy is done by using the LaB$_{6}$ NIST standard sample and wavelength of the x-ray is estimated to be $0.80471$~\AA. Rietveld refinement of all the XRD data is performed using FullProf Software Package.\cite{Rodriguez55}

The DC magnetization ($M$) measurements as a function of temperature and magnetic field ($H$) are performed using two different magnetometers: Vibrating Sample Magnetometer (VSM) option of 9~T PPMS and 7~T SQUID magnetometer, all from M/s. Quantum Design, USA. For each measurement, the magnetic field is lowered to zero from a high field value in the oscillating mode at high temperatures (above the magnetic transition) in order to minimize the residual field. For the magnetic isotherms (at and below $T_{\rm C}$), the demagnetization field ($H_{\rm dem}$) has been subtracted from the applied field ($H_{\rm ex}$) following the procedure described in Ref.~\cite{Kaul1114}. The temperature dependent resistivity measurements (3-300~K) are performed using four probe method in a home-made resistivity set-up attached to a cryostat (M/s. OXFORD Instrument, UK) with 8~T superconducting magnet.

\section*{Results and Discussion}
\begin{figure}[htbp]
	\centering
	\includegraphics[width = 7.5 cm]{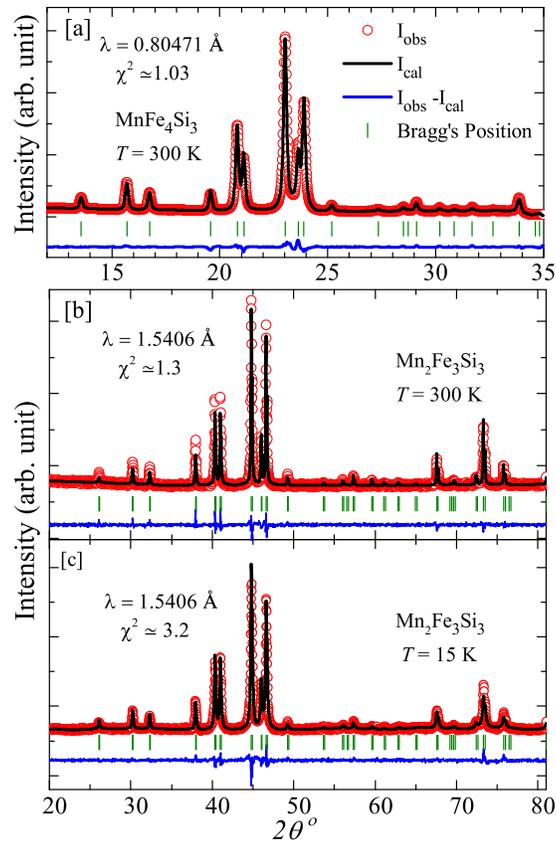}
	\caption{\label{Fig1} The powder XRD patterns of [a] MnFe$_{4}$Si$_{3}$ at room temperature using the synchrotron source, [b] Mn$_{2}$Fe$_{3}$Si$_{3}$ at room temperature using the lab source, and [c] Mn$_{2}$Fe$_{3}$Si$_{3}$ at $T = 15$~K using the lab source. The solid line represents the Rietveld refinement of the experimental data, the green vertical bars correspond to Bragg positions, and the bottom blue line represents the difference between observed and calculated intensities.}
\end{figure}
\begin{figure}[htbp]
	\centering
	\includegraphics[width = 7.5 cm,height = 8.5 cm]{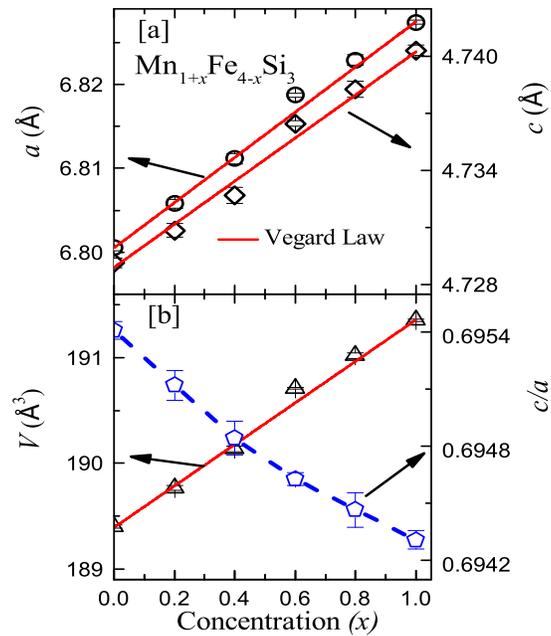}
	\caption{\label{Fig2} The variations of lattice parameters [a] $a$ and $c$ and [b] unit cell volume $V$ and $c/a$ ratio as a function of Mn concentration ($x$). The solid lines are the fits using Vegard's law, as described in the text.}
\end{figure}
\begin{figure}[htbp]
	\centering
	\includegraphics[width = 7.5 cm,height = 8 cm]{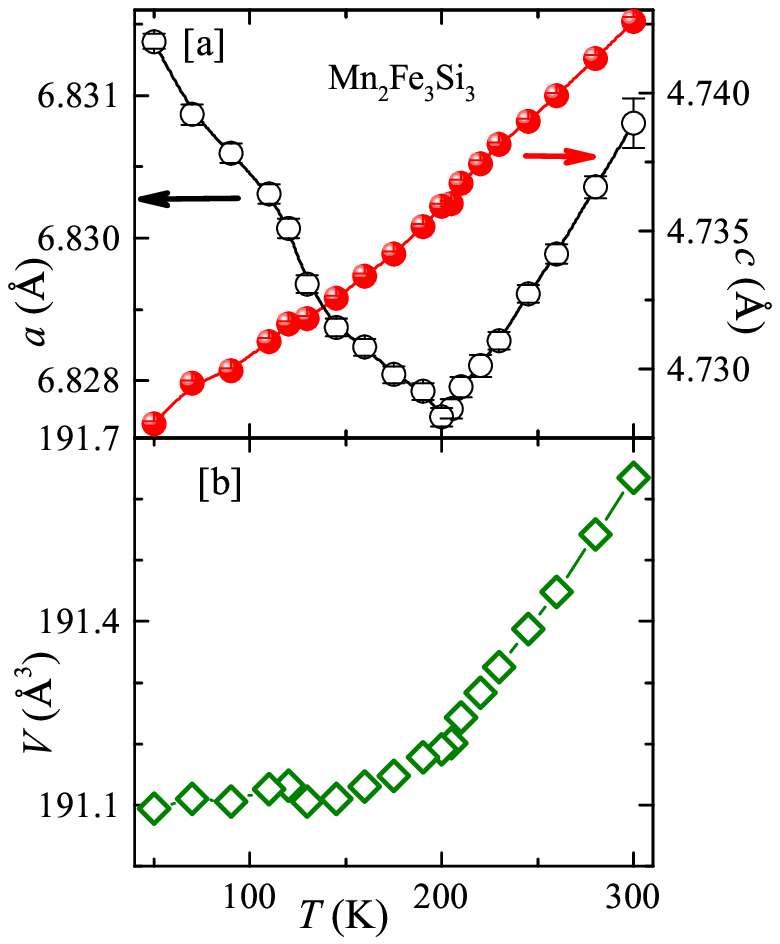}
	\caption{\label{Fig3} Temperature variation of lattice parameters [a] $a$ and $c$ and [b] unit cell volume $V$, obtained from the Rietveld refinement of the powder XRD patterns of Mn$_{2}$Fe$_{3}$Si$_{3}$.}
\end{figure}
\subsection*{X-ray Diffraction}
Figure~\ref{Fig1}[a] presents the room temperature powder XRD pattern of MnFe$_{4}$Si$_{3}$ measured at the synchrotron facility. Clearly, our synchrotron data do not show any extra peak associated with the foreign phases and all the peaks could be indexed using hexagonal crystal structure with space group $P6_{3}/mcm$.\cite{Johnson311} Our Rietveld analysis also confirms that the sample is single phase with the lattice parameters $a = 6.8070(4)$~\AA, $c = 4.7341(3)$~\AA, and unit cell volume $V = 189.97(2)$~\AA$^{3}$ which are in good agreement with the previous reports.\cite{Binczycka_K13,Johnson311,Hering7128,Gourdon56}
The XRD patterns for other compositions ($x=$ 0.2, 0.4, 0.6, 0.8, and 1.0) indicate that Mn substitution at the Fe site does not alter the symmetry of the crystal structure, but shifts the major XRD peaks to lower $2\theta$ values. A representative XRD pattern at room temperature with Rietveld refinement is shown for the end composition ($x=1.0$) in Fig.~\ref{Fig1}[b]. The obtained lattice parameters for $x=1$ are also in good agreement with the reported values.\cite{Binczycka_K13,Johnson311}
The variation of lattice parameters ($a$, $c$, and $V$) with $x$ is presented in Fig.~\ref{Fig2}. It shows that $a$, $c$, and $V$ increase linearly with $x$ which can be fitted nicely using Vegard's law.\cite{Vegard3161} This suggests that Mn replaces Fe in the unit cell, leading to a lattice expansion since Mn has larger atomic radius than Fe. The $M1$ atoms with Wyckoff position $4d$ make a chain along the $c$-axis whereas the $M2$ atoms with Wyckoff position $6g$ are surrounded by two other $M2$ atoms in the plane perpendicular to the $c$-axis.\cite{Johnson311} The almost linear decreases of $c/a$ with increasing $x$ suggests that the expansion of the unit cell is more along the $a$-direction compared to the $c$-direction. This also further indicates that Mn preferentially replaces Fe at the $6g$ site in the crystal lattice.

From the temperature dependent XRD and neutron diffraction studies a change of slope in $V(T)$ and a minima in $a(T)$ are reported for the parent compound MnFe$_{4}$Si$_{3}$ across the PM-FM transition ($T_{\rm C} \simeq 300$~K), without altering the crystal symmetry.\cite{Gourdon56,Hering7128} In order to check how the Mn substitution affects this feature, temperature dependent XRD measurements are performed on the end composition Mn$_{2}$Fe$_{3}$Si$_{3}$ ($x = 1.0$). Figure~\ref{Fig1}[c] presents the XRD pattern along with the Rietveld refinement at 15~K. The crystal structure for $x=1.0$ remains unchanged down to 15~K, similar to the parent compound. The temperature variation of $a$, $c$, and $V$ are shown in Fig.~\ref{Fig3}. With increasing $T$, $c$ increases monotonically while $a$ decreases, resulting in a nearly constant unit cell volume up to 200~K which corresponds to the FM transition temperature. Above 200 K or in the PM state, both $a$ and $c$ increase linearly with $T$, as a consequence, $V$ also increases linearly with $T$.

\begin{figure}[htbp]
	\centering
	\includegraphics[width = 8 cm]{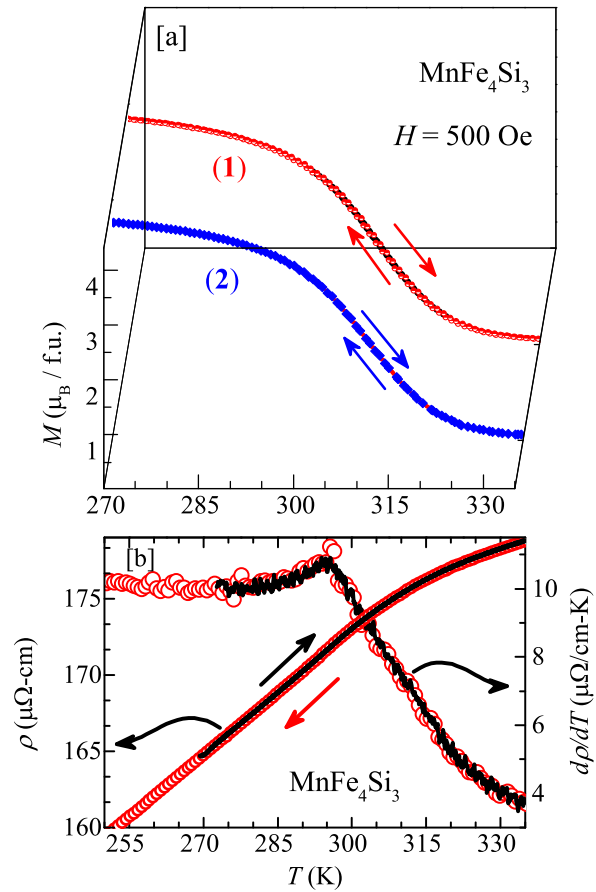}
	\caption{\label{Fig4} [a] Magnetization as a function of temperature at $H=500$~Oe for MnFe$_{4}$Si$_{3}$ sample measured during cooling and warming, using VSM and SQUID magnetometers are denoted as: (1) VSM with 1~K/min in settle mode and (2) SQUID magnetometer with 1~K/min in sweep mode. [b] Zero field resistivity ($\rho$) measured during cooling and warming cycles and its derivative ($d\rho/dT$) vs $T$ are plotted along the left and right $y$-axes, respectively.}
\end{figure}
\subsection*{PM-FM Transition} 
Magnetization ($M$) as a function of temperature for the parent compound MnFe$_{4}$Si$_{3}$ measured in an applied field of $H=500$~Oe, during cooling and warming is presented in Fig.~\ref{Fig4}[a]. Measurements are done using both VSM and SQUID magnetometers. The rapid increase in $M$ around $310$~K indicates the PM to FM transition, consistent with the previous reports.\cite{Gourdon56,Hering7128} Previously, Hering~et~al observed a thermal hysteresis across the magnetic transition which was taken as a signature of the first order PM-FM phase transition.\cite{Hering7128} Our measurements using VSM in temperature sweep mode during cooling and warming exhibits a large thermal hysteresis ($\sim 3$~K) across the magnetic transition (not shown). On the other hand, when the measurements are done using the same VSM in the settle mode (i.e. after stabilizing at each temperature) (labeled as \textbf{1}), the hysteresis is reduced substantially ($\sim 0.7$~K). To further check the hysteresis behaviour, $M$ vs $T$ was measured using SQUID magnetometer (labeled as \textbf{2}). As shown in Fig.~\ref{Fig4}[a], the measurements during cooling and warming show almost no hysteresis. Furthermore, temperature dependent resistivity [$\rho(T)$] measurement also does not show any signature of thermal hysteresis during cooling and warming (see Fig.~\ref{Fig4}[b]). The temperature derivative of resistivity [$d\rho/dT$] as a function of $T$ is also shown in the same figure to highlight the transition and no hysteresis. These results demonstrate that the thermal hysteresis reported by Hering~et~al could be a measurement artifact.\cite{Hering7128}
\begin{figure}[htbp]
	\centering
	\includegraphics[width = 8 cm,height = 11 cm]{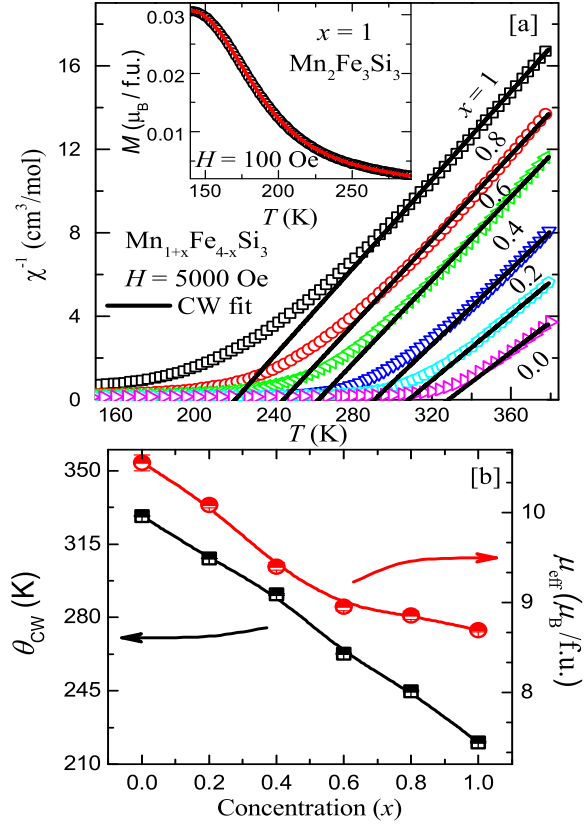}
	\caption{\label{Fig5} [a] Temperature dependent inverse susceptibility $\chi^{-1}$ of Mn$_{1+x}$Fe$_{4-x}$Si$_{3}$ for $ x = 0.0, 0.2, 0.4, 0.6, 0.8 $, and $ 1.0 $. The solid lines are the CW fits. Inset: Magnetization as a function of temperature measured during cooling and heating for the end composition $x=1$. [b] Variation of CW temperature $\Theta_{\rm CW}$ and effective magnetic moment $ \mu_{\rm eff} $ with the doping concentration ($x$).}
\end{figure}

It is also predicted that one would observe the Landau tricritical point in the vicinity of the PM-FM transition in the parent MnFe$_{4}$Si$_{3}$ compound.\cite{Herlitschke094304} Therefore, we tried to tune the PM-FM transition to lower temperatures by Mn substitution at the Fe site. Figure~\ref{Fig5}[a] presents the temperature dependent inverse susceptibility $\chi^{-1}$ [$\equiv (M/H)^{{-1}}$] measured at $H=5000$~Oe for Mn$_{1+x}$Fe$_{4-x}$Si$_{3}$ with $x = 0.0, 0.2, 0.4, 0.6, 0.8$, and 1. It shows that the PM to FM transition shifts to low temperatures with increasing $x$. Each curve in the high temperature range (well above $T_{\rm C}$) is fitted using Curie-Weiss (CW) law
\begin{equation}\label{eq.1}
	\chi(T)=\dfrac{C}{{T-\Theta_{\rm CW}}},
\end{equation}
where, $C$ is the Curie constant and $\Theta_{\rm CW}$ is the CW temperature. For the parent compound, the CW fit provides $\Theta_{\rm CW} \simeq 328.3$~K and the effective magnetic moment $\mu_{\rm eff} \simeq 2.11(1)$~$\mu_{\rm B}/$transition metal atom. These values are in close agreement with the previous reports.\cite{Hering7128,Herlitschke094304} 
The CW fits show that $\Theta_{\rm CW}$ is shifting systematically towards low temperatures with increasing Mn concentration as shown in Fig.~\ref{Fig5}[a].
The obtained $\Theta_{\rm CW}$ and $\mu_{\rm eff}$ are plotted as a function of $x$ in the left and right $y$-axes, respectively in Fig.~\ref{Fig5}[b]. Both the parameters decrease systematically with increasing $x$.\cite{Herlitschke094304,Johnson311} An almost linear decrease of $\Theta_{\rm CW}$ reflects the effect of dilution which apparently tunes the exchange energy. Thus, as the Mn concentration increases, the unit cell volume increases which weakens the exchange interaction. Moreover, the electronic contribution due to Mn substitution at the Fe site can also be partly responsible for the variation of $\theta_{\rm CW}$ with $x$, which cannot be completely ignored in the present study.
Further, no thermal hysteresis across the PM-FM transition is observed for any compositions even in a very low field of 10~Oe. A representative magnetization curve taken during cooling and warming in $H = 100$~Oe for the end composition $x=1.0$ is shown in the inset of Fig.~\ref{Fig5}[a] indicating second-order nature of the transition. This also rules out the possibility of a tricritical point, opposing the previous prediction.\cite{Herlitschke094304}

\subsection*{Critical Scaling}
The critical analysis of the magnetization data were carried out for the compositions $x = 0$ and 0.2 following the procedure described in Refs.~\cite{Fischer064443,Pramanik214426}. The critical or scaling analysis is typically carried out by measuring magnetization isotherms ($M$ vs $H$) in the vicinity of $T_{\rm C}$ for a second order ferro/ferri-magnetic transition which provides information about the universality class of the system. The set of critical exponents ($\beta$, $\gamma$, and $\delta$) characterizing the phase transition can be obtained from the analysis of the spontaneous magnetization ($M_{\rm S}$), zero field susceptibility ($\chi_{0}$), and magnetization isotherm at the $T_{\rm C}$, following the set of relations (Power Laws)\cite{Stanley336}
\begin{equation}\label{eq.2}
  \centering
	M_{\rm S}(T) = M_{0}(-\epsilon)^{\beta},~{\rm for}~\epsilon < 0, T<T_{\rm C},
\end{equation}
\begin{equation}\label{eq.3}
	\chi_0 ^{-1}(T) = \Gamma(\epsilon)^{\gamma},~{\rm for}~\epsilon > 0,  T > T_{\rm C},
\end{equation}
\begin{equation}\label{eq.4}
	M(H) = X(H)^{1/\delta},~{\rm for}~\epsilon = 0, T = T_{\rm C}.
\end{equation}
Here, $\epsilon = \dfrac{T-T_{\rm C}}{T_{\rm C}}$ is the reduced temperature and $M_{0}$, $\Gamma$, and $X$ are the critical coefficients. These critical exponents are related to each other as
\begin{equation}\label{eq.5}
	\delta = 1 + \frac{\gamma}{\beta}.
\end{equation} 
These exponents also satisfy the following equation of state which relates magnetization $M$ with $H$ and $T$
\begin{equation} \label{eq.6}
	M(H,\epsilon)\left\lvert\epsilon\right\rvert^{-\beta}=\textit{f}_{\pm}(H\left\lvert\epsilon\right\rvert^{-(\beta +\gamma)}).
\end{equation}
Here, $\textit{f}_{+}$ and $\textit{f}_{-}$ are the scaling functions above and below $T_{\rm C}$, respectively. The renormalization of scaling [Eq.~(\ref{eq.6})] in term of reduced magnetization $m = M(H,\epsilon)\epsilon^{-\beta}$ and reduced susceptibility $h/m = (H/M)\epsilon^{-\gamma}$ leads to a much sensitive equation of state\cite{Fischer064443}
\begin{equation}
	h/m = \pm a_{\pm} + b_{\pm} m^{2}.
	\label{eq.7}
\end{equation}
Here, $+$ and $-$ correspond to the temperatures above and below $T_{\rm C}$, respectively. With the appropriate values of $\beta$, $\gamma$, and $T_{\rm C}$, the curves obtained from the implementation of both the equations [Eq.~(\ref{eq.6}) and Eq.~(\ref{eq.7})] will collapse into two separate universal branches: one above and another below the $T_{\rm C}$.

\subsubsection*{Arrott Plot}
\begin{figure}[htbp]
	\centering
	\includegraphics[width=16cm]{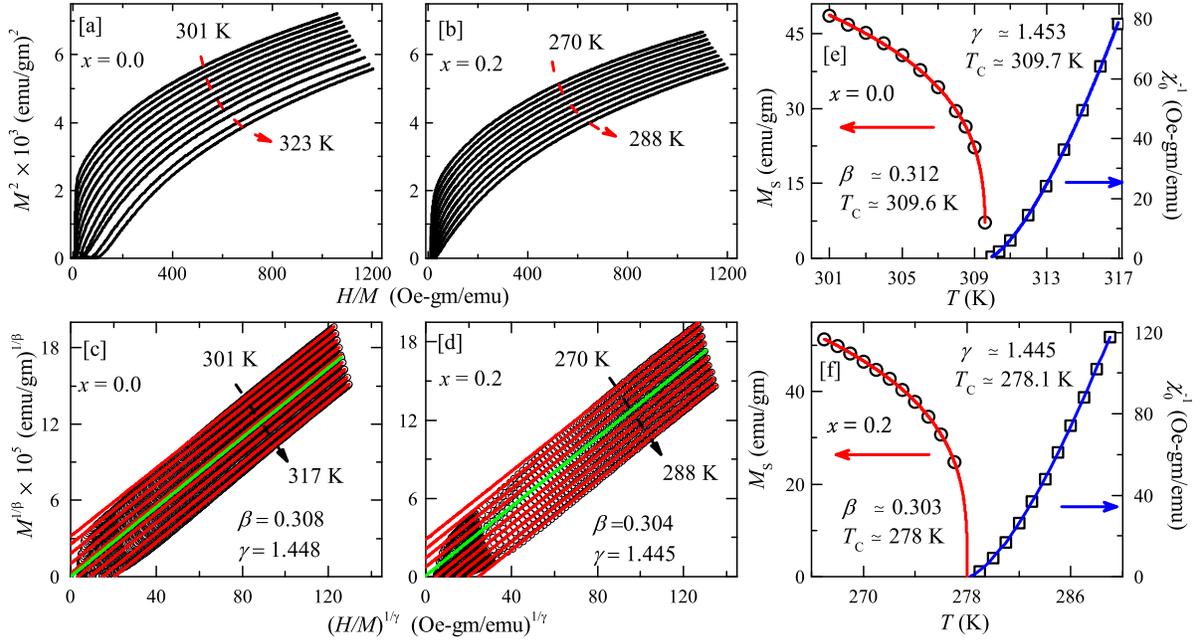}
	\caption{\label{Fig6} The Arrott plots ($M^{2}$ vs $H/M$) for [a] MnFe$_{4}$Si$_{3}$ ($x=0$) and [b] Mn$ _{1.2}$Fe$ _{3.8}$Si$_{3}$ ($x = 0.2$) at various temperatures, above and below $T_{\rm C}$. The modified Arrott plots ($M^{1/\beta}$ vs $H/M^{1/\gamma}$) for [c] $x=0.0$ and [d] $x=0.2$. The solid lines are the linear fits to the data in the high field regime ($H \geq 2.5$~T) and are extrapolated to $H/M = 0$. Spontaneous magnetization $M_{\rm S}$ and zero field inverse susceptibility $\chi_{0} ^{-1}$ as a function of temperature in the left and right $y$-axes, respectively for [e] $x = 0.0$ and [f] $x=0.2$, obtained from the intercepts of the modified Arrott plots in the vicinity of $T_{\rm C}$. The solid lines are the fits as described in the text.}
\end{figure}

Arrott plot is a very useful and standard method for establishing the onset of ferromagnetic/ferrimagnetic transition and also for an accurate determination of $T_{\rm C}$ and critical exponents.\cite{Arrott1394} According to the mean field theory, the $M^{2}$ vs $H/M$ plots should be straight and parallel lines and the curve at the $T_{\rm C}$ should pass through origin. However, experimentally such Arrott plots can exhibit considerable curvature arising from the non mean-field type behaviour. Therefore, modified Arrott plots (MAP) are used where $M^{1/\beta}$ is plotted against $(H/M)^{1/\gamma}$.\cite{Arrott786} From the values of the critical exponents ($\beta$ and $\gamma$) that give straight line curves, the universality class of the spin system is uniquely decided. The Arrott plots ($M^{2}$ vs $H/M$) constructed out of the magnetization isotherms in the vicinity of $T_{\rm C}$ are shown in Fig.~\ref{Fig6}[a] and [b] for two compositions $x= 0.0$ and 0.2, respectively. Clearly, in our case, the $M^{2}$ vs $H/M$ plots deviate from the straight line behavior suggesting that the mean-field model is inadequate to explain the transition. Moreover, according to the Banerjee criterion, the positive slope of the $M^{2}$ vs $H/M$ curves indicates the second order nature of the PM to FM transition for both the samples.\cite{Banerjee16}

Next, we used the modified Arrott plots (MAP) based on the Arrott-Noakes equation.\cite{Arrott786} In order to obtain the acceptable values of $\beta$ and $\gamma$, we have followed the iterative method described in Refs.~\cite{Fischer064443,Pramanik214426} and the starting trial values are taken to be $\beta = 0.365$ and $\gamma = 1.386$, corresponding to the 3D Heisenberg model. Using these values of $\beta$ and $\gamma$, initial MAPs are obtained from the magnetic isotherms at different temperatures, around $T_{\rm C}$. The data in the high field regime of the MAPs are fitted by a straight line and are extrapolated to obtain the spontaneous magnetization [$M_{\rm S}(T)$] and inverse of the zero field susceptibility [$\chi_{0} ^{-1}(T)$] from the intercepts on the $M^{1/\beta}$ and $(H/M)^{1/\gamma}$ axes, respectively. These values of $M_{\rm S}(T)$ and $\chi_{0}^{-1}(T)$ are further fitted using Eq.~(\ref{eq.2}) and (\ref{eq.3}), respectively to obtain a more reliable set of $\beta$, $\gamma$, and $T_{\rm C}$ values. These new set of $\beta$ and $\gamma$ are again used to construct another set of MAPs. This procedure was carried out for few iterations after which a set of stable values of $\beta$, $\gamma$, and $T_{\rm C}$ are arrived and MAPs are found to be straight lines. The final MAPs are presented in Fig.~\ref{Fig6}[c] and [d] with ($\beta \simeq 0.308$, $\gamma \simeq 1.448$) and ($\beta \simeq 0.304$, $\gamma \simeq 1.445$) for $x = 0.0$ and $0.2$, respectively. Similarly, the final $M_{\rm S}$ and $\chi_{0}^{-1}$ as a function of temperature, below and above $T_{\rm C}$ are plotted in Fig.~\ref{Fig6}[e] and [f] for $x=0.0$ and $x=0.2$, respectively.
\begin{figure}[htbp]
	\centering
	\includegraphics[width = 7 cm]{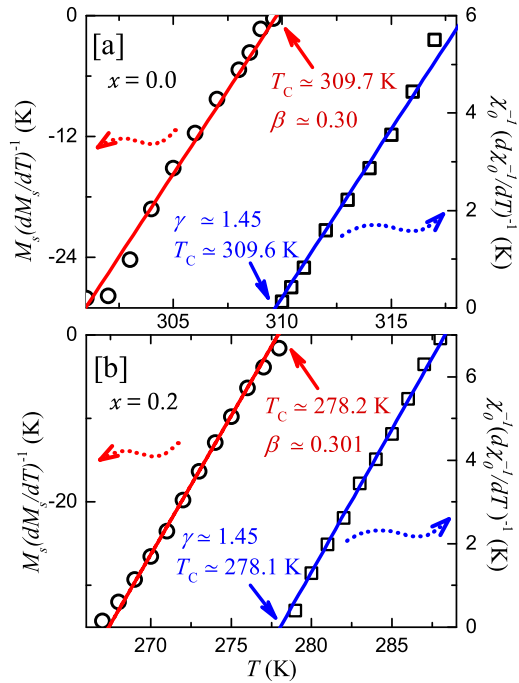}
	\caption{\label{Fig7} The Kouvel-Fisher plot of $ M_{\rm S} $ and $ \chi_0^{-1}$ for [a] $x=0.0$ and [b] $x=0.2$. The solid lines are the linear fits. The solid arrows point to the $T_{\rm C}$s.}
\end{figure}
\begin{figure}[htbp]
	\centering
	\includegraphics[width = 7 cm]{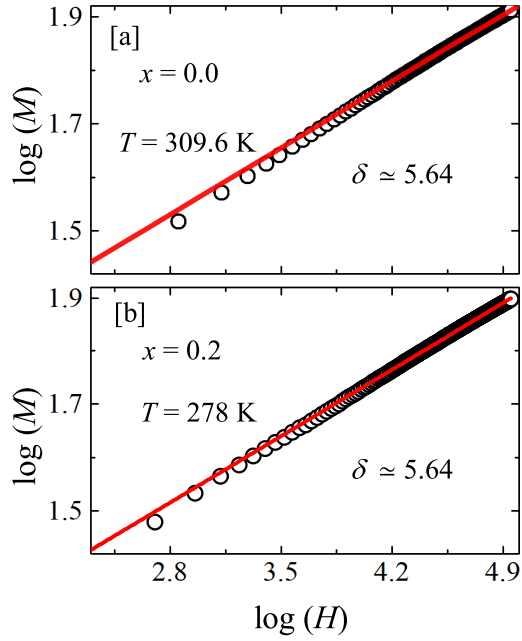}
	\caption{\label{Fig8} The log-log plot of isothermal magnetization ($M$) vs applied field ($H$) for [a] $x=0.0$ and [b] $x=0.2$ collected at their respective $T_{\rm C}$s. The solid lines are the linear fits of critical isotherm relation described in Eq.~(\ref{eq.4}) to extract the $\delta$ value.}
\end{figure}
These final $M_{\rm S}(T)$ and $\chi_{0}^{-1}(T)$ data are fitted using Eq.~(\ref{eq.2}) and (\ref{eq.3}), respectively. The obtained critical parameters and $T_{\rm C}$s are [($\beta \simeq 0.312$ and $T_{\rm C} \simeq 309.6$~K) from $M_{\rm S}$ and ($\gamma \simeq 1.453$ and $T_{\rm C} \simeq 309.7$~K) from $1/\chi_0$] and [($\beta \simeq 0.303$ and $T_{\rm C} \simeq 278.01$~K) from $M_{\rm S}$ and ($\gamma \simeq 1.445$ and $T_{\rm C} \simeq 278.14$~K) from $1/\chi_0$] for $x=0.0$ and 0.2, respectively. All these exponents and $T_{\rm C}$ values are summarized in Table.~\ref{arttype}. The estimated values of $\beta$, $\gamma$, and $T_{\rm C}$ using Eq.~(\ref{eq.2}) and (\ref{eq.3}) are very close (within error bars) to the values obtained from the MAPs in Fig.~\ref{Fig6}[c] and [d].

\subsubsection*{Kouvel-Fisher Plot}
The values of $\beta$, $\gamma$, and $T_{\rm C}$ can further be estimated more reliably by analyzing the $M_{\rm S}(T)$ and $\chi_{0}^{-1}(T)$ data, obtained from the MAPs, in terms of the Kouvel-Fisher plots (KFPs).\cite{KouvelA1626} In this method, $M_{\rm S}(T)(dM_{\rm S}(T)/dT)^{-1}$ and $\chi_0 ^{-1}(T)(d\chi_0 ^{-1}(T)/dT)^{-1}$ are plotted as a function of temperature which are expected to produce straight line curves. When fitted by a straight line, the $x$-intercepts give value of $T_{\rm C}$ and the inverse of the slopes provides the value of critical exponents ($\beta$ and $\gamma$), respectively. As shown in Fig.~\ref{Fig7}, a linear fit to the data results [($\beta \simeq 0.30$ and $T_{\rm C} \simeq 309.7$~K) from $M_{\rm S}$ and ($\gamma \simeq 1.45$ and $T_{\rm C} \simeq 309.6$~K) from $\chi_0^{-1}$] and [($\beta \simeq 0.301$ and $T_{\rm C} \simeq 278.2$~K) from $M_{\rm S}$ and ($\gamma \simeq 1.45$ and $T_{\rm C} \simeq 278.1$~K) from $\chi_0^{-1}$] for $x=0.0$ and 0.2, respectively. These values of $\beta$, $\gamma$, and $T_{\rm C}$ are found to be quite consistent with the ones obtained from the MAP analysis.

\subsubsection*{Critical Isotherm}
To extract another critical exponent $\delta$ as given in Eq.~(\ref{eq.4}), one can plot $\log(M)$ vs $\log(H)$ of the critical magnetization isotherm at the $T_{\rm C}$. The reciprocal of the slope of a linear fit would provide the value of $\delta$. As depicted in Fig.~\ref{Fig8}, our $\log(M)$ vs $\log(H)$ plot at the $T_{\rm C}$ (i.e. at $T_{\rm C} \simeq 309.6$~K for $x=0.0$ and $T_{\rm C} \simeq 278$~K for $x=0.2$) is almost linear. A straight line fit over the whole measured field range results the same value of $\delta \simeq 5.64$ for both the compositions. Furthermore, $\delta$ can also be calculated using the Widom scaling relation $\delta  =  1 + \frac{\gamma}{\beta}$ where two of the three exponents are independent.\cite{Widom1633,Widom3898} Using the appropriate values of $\beta$ and $\gamma$, obtained from the MAPs we found $\delta \simeq 5.70$ for both the compounds which matches well with the value obtained above from the critical isotherm at the $T_{\rm C}$. This further confirms the self-consistency of our estimation of critical exponents.
\begin{figure}[htbp]
	\centering
	\includegraphics[width = 7 cm, height = 10 cm]{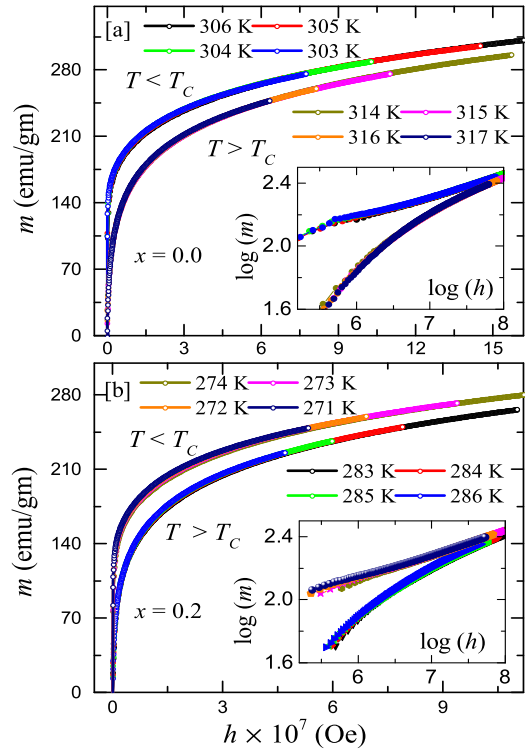}
	\caption{\label{Fig9} The reduced magnetization $(m = M\mid\epsilon\mid^{-\beta}) $ vs reduced magnetic field $(h=H\mid\epsilon\mid^{-(\gamma+\beta)})$ plot for [a] $x=0.0$ and [b] $x=0.2$. The renormalized curves in different temperatures just above and below $T_{\rm C}$ are collapsing into two separate branches. Inset: log($m$) vs log($h$) to magnify the low field dispersions.}	     
\end{figure}
\begin{figure}[htbp]
	\centering
	\includegraphics[width = 7 cm, height = 10 cm]{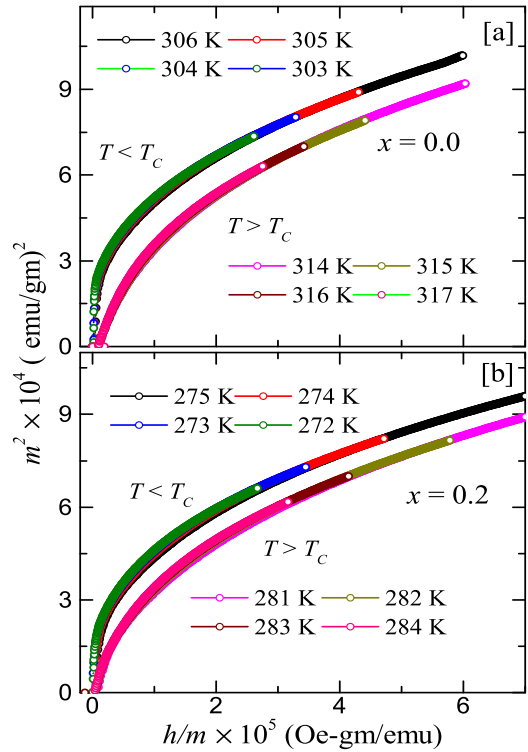}
	\caption{\label{Fig10} The $m^{2}$ vs $h/m$ plot for [a] $x=0$ and [b] $x=0.2$ taking the $m$ and $h$ data from Fig.~\ref{Fig9}. The renormalized curves in [a] and [b] at different temperatures just above and below $T_{\rm C}$ are collapsing into two separate branches.}	     
\end{figure}
\begin{figure}[htbp]
	\centering
	\includegraphics[width = 7 cm, height = 10 cm]{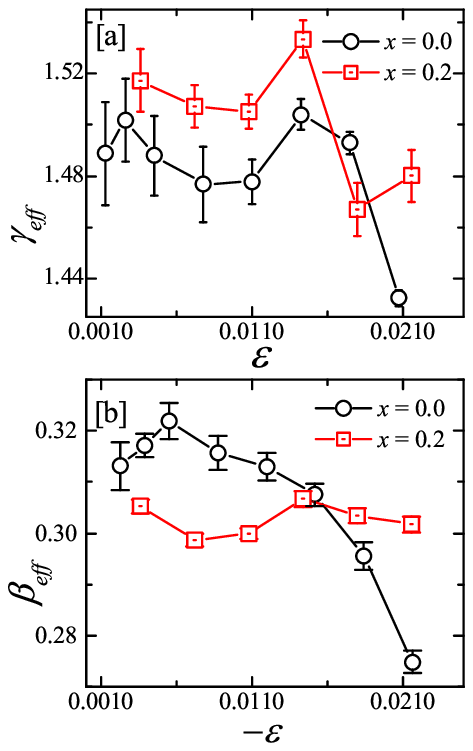}
	\caption{\label{Fig11} The effective critical exponents [a] $\gamma_{eff}$ and [b] $\beta_{eff}$ for $x=0.0$ and 0.2 samples are plotted as a function of reduced temperature $\epsilon$, above and below $T_{\rm C}$.}	     
\end{figure}

\subsubsection*{Validity of Scaling Law}
The values of critical exponents estimated via different methods are tabulated in Table~\ref{arttype}. The theoretically expected values for mean-field model, 3D Heisenberg model, and 3D Ising model are also listed for a comparison. The values of critical exponents for $x = 0$ and $0.2$ are found to be almost same, reflecting similar kind of interaction in both the systems. The same analysis, is likewise, done for the $x=0.4$ sample (not shown) and the value of the critical exponents are found to be identical to that of $x=0.0$ and $x=0.2$ samples with $T_{\rm C} \simeq 253.8$~K. It is interesting to note that our experimental values of critical exponents do not coincide with any of the standard universality classes. Moreover, these critical exponents also do not match with other reported compounds in the literatures. The most closest critical exponents are found to be ($\beta\sim 0.30$, $\gamma\sim 1.39$, and $\delta \sim 5.5$) and ($\beta\sim 0.315$, $\gamma\sim 1.39$,$\delta \sim 5.36$) corresponding to Cr$_{75}$Fe$_{25}$ and Cr$_{70}$Fe$_{30}$, respectively.\cite{Fischer064443} Hence, to further check the reliability of the critical exponents we attempted to generate the scaling equation [Eq.~(\ref{eq.6})] using these values. For this purpose, we renormalized the isotherms following Eq.~(\ref{eq.6}) and using final values of $\beta$, $\gamma$, and $T_{\rm C}$ from MAP analysis (Table~\ref{arttype}). Figure~\ref{Fig9}[a] and [b] present the reduced magnetization ($m$) vs the reduced field ($h$) for $x=0.0$ and 0.2, respectively. Here, we have chosen four temperatures above and four temperatures below the $T_{\rm C}$. Clearly, these curves collapse into two separate branches in which the isotherms just above $T_{\rm C}$ form the lower curve and the isotherms just below $T_{\rm C}$ form the upper curve in Fig.~\ref{Fig9}. We have also plotted log($m$) vs log($h$) in the insets in order to highlight the two branches and no deviations in the low field regime. Another robust method to ensure the reliability of $\beta$, $\gamma$, and $T_{\rm C}$ is to plot $\textit{m}^{2}$ vs $\textit{m/h}$ for temperatures just above and below the $T_{\rm C}$ following Eq.~(\ref{eq.7}). As reflected in Fig.~\ref{Fig10}, all the isotherms collapse into two separate branches: one above the $T_{\rm C}$ and another below the $T_{\rm C}$. The above analysis confirms the reliability of the critical exponents and suggests that the interactions get renormalized at the critical regime following the equation of state.

\subsubsection*{Effective Critical Exponents}
Our estimated critical exponents do not fall in any of the common universality classes. Often the exponents are strongly influenced by various factors such as competing interactions, disorder etc. However, the real exponents reflecting the true universality class of the compounds can be assessed by performing the analysis only in the critical regimes when $\epsilon \rightarrow 0$. Therefore, it is interesting to check what happens to these exponents while approaching the asymptotic/critical limit. We calculated the effective critical exponents ($\beta_{eff}$ and $\gamma_{eff}$) from the analysis of $M_{\rm S}$ and $\chi_{0}^{-1}$, respectively just above and below the $T_{\rm C}$, using the equations\cite{Pramanik214426}
\begin{equation}
	\beta_{\rm eff}(\epsilon) = \dfrac{d[lnM_{S}(\epsilon)]}{{d(ln\epsilon)}},~~~~~\gamma_{\rm eff}(\epsilon) = \dfrac{d[ln\chi_{0}^{-1}(\epsilon)]}{{d(ln\epsilon)}}.
	\label{eq.8}
\end{equation}
The obtained values of $\gamma_{\rm eff}$ and $\beta_{\rm eff}$ are plotted as a function of reduced temperature $\epsilon$ in Fig.~\ref{Fig11}[a] and [b], respectively for $x = 0.0$ and 0.2. For both the compounds $\beta_{\rm eff}$ and $\gamma_{\rm eff}$ show a nonmonotonic change with $\epsilon$ and approach a value of 0.31 and 1.5, respectively at the lowest investigated $\epsilon$ of $\sim 10^{-3}$. These values are much closer to the critical exponents $\beta$ and $\gamma$ listed in Table~\ref{arttype}, obtained from various analysis schemes and they seem to converse to the actual values in the asymptotic regime ($\epsilon \rightarrow 0$). This further reflects not only that the compounds under investigation do not fall in any of the known universality classes and but also our analysis is complete in all respect.

\subsubsection*{Spin Interaction}
The universality class of the phase transition depends on the nature of exchange interaction. According to the renormalization group theory, the isotropic interaction $J(r)$ in $d$-dimensions decays following\cite{Fisher917} 
\begin{equation}\label{eq.9}
	J(r) \sim r^{-\rm (d+\sigma)},
\end{equation}
where, $\sigma$ is a positive constant which represents the range of interaction and $r$ is the distance.
In this model, $\sigma < 2$ implies long range interaction while $\sigma > 2$ reflects short range interaction. From the value of $\sigma$, the critical exponent $\gamma$ can be estimated theoretically as\cite{Fisher917}
\begin{equation}\label{eq.10}
\gamma = 1 +\dfrac{4}{d}\Big(\dfrac{n+2}{n+8}\Big)\Delta\sigma +\dfrac{8(n+2)(n-4)}{d^{2}(n+8)^{2}}\times\left[1 +\dfrac{2G(\frac{d}{2})(7n+20))}{(n-4)(n+8)}\right]\Delta\sigma^{2},
\end{equation}
where, $\Delta \sigma = (\sigma-\frac{d}{2})$, $ G(\frac{d}{2}) = 3-\frac{1}{4}(\frac{d}{2})^{2}$, and $d$ and $n$ are the lattice dimensionality and spin dimensionality, respectively. Here, one needs to choose the value $\sigma$ in Eq.~(\ref{eq.10}) in such a way that a particular set of $d$ and $n$ values should yield a $\gamma$ value close to the experimental one. Using the value of $\sigma$, other critical exponents can further be calculated as $\nu = \gamma/\sigma, \eta = 2-\sigma, \alpha = 2-\nu d, \beta = (2-\alpha-\gamma)/2$, and $\delta = 1 + \gamma/\beta$.\cite{Fisher917,Fischer064443} The choice of $(d:n)= (2:1)$ and $\sigma = 1.41$ produce $\gamma = 1.445$, which is close to our experimentally observed value ($\sim 1.45$). This implies long-range spin-spin interaction in the system under investigation. Using the values $\sigma \simeq 1.41$, $d = 2$, and $n =1$, the other critical exponents are estimated to be $\beta \simeq 0.300$, $\gamma \simeq 1.448$, $\delta \simeq 5.831$, $\nu \simeq 1.02$, $\eta \simeq0.586$, and $\alpha \simeq-0.0475$. These values are quite consistent with the values obtained from other methods as listed in Table~\ref{arttype}. Thus, the exchange interaction between magnetic spins decays with distance as $J(r)\sim r^{-3.41}$. Indeed, our findings are quite identical to that reported for Cr$_{75}$Fe$_{25}$ and Cr$_{70}$Fe$_{30}$ where the value of critical exponents coincide with the ones calculated from the renormalization group theory for $d=2$ and $n=1$ with a long-range interaction between the spins.\cite{Fischer064443}
\begin{table}[htbp]
	\centering
	\begin{tabular}{|l|l|l|l|l|l|l|}
			\hline
			System&$\beta$&$\gamma$&$\delta$&$T_{\rm C}(K)$&Method &Refs  \\
			\hline
			$x =0.0$&0.308(3)&1.448(5)&5.641(4)&309.60(2)&MAP & \\ 
			&0.303(4)&1.451(4)&5.77(7)&309.7(1)&KF & This work \\
			&--&--&5.644(9)&309.6&Critical Isotherm & \\ 
			&--&--&5.70(7)&309.6&MCE/RCP & \\
			&--&--& 5.70 & -- & Widom scaling & \\
			\hline 
			$x = 0.2$ & 0.304 & 1.445 & 5.75 & 278.17(3) & MAP & \\ 
			& 0.301(1) & 1.45(1) & 5.77(4)& 278.1(1) & KF & This work \\
			&--&--& 5.64(3) & 278 & Critical Isotherm & \\ 
			&--&--& 5.73(13) & 278 & MCE/RCP & \\
			&--&--& 5.70 & -- & Widom scaling & \\
			\hline 
			Mean Field Model & 0.5& 1.0& 3.0& --& &\cite{Kaul5}\\
			\hline 
			3D Heisenberg Model &0.365 &1.386 & 4.80&-- & &\cite{Kaul5}\\
			\hline 
			3D Ising Model &0.325 & 1.241& 4.82& && \cite{Kaul5}\\
			\hline 
		\end{tabular}
		\caption{\label{arttype} The obtained values of critical exponents ($\beta$, $\gamma$, and $\delta$) and $T_{\rm C}$s from the modified Arrott plot (MAP), Kouvel-Fisher (KF) plot, critical isotherm, Widom scaling, and magnetocaloric effect (MCE)/relative cooling power (RCP) analysis across the PM-FM transition ($T_{\rm C}$) for Mn$_{1+x}$Fe$ _{4-x}$Si$_{3}$ ($x=0$ and 0.2). For completeness, we have also tabulated the theoretically predicted values of the critical exponents for different universality classes.}
\end{table}

\begin{figure}[htbp]
	\centering
	\includegraphics[width=18 cm]{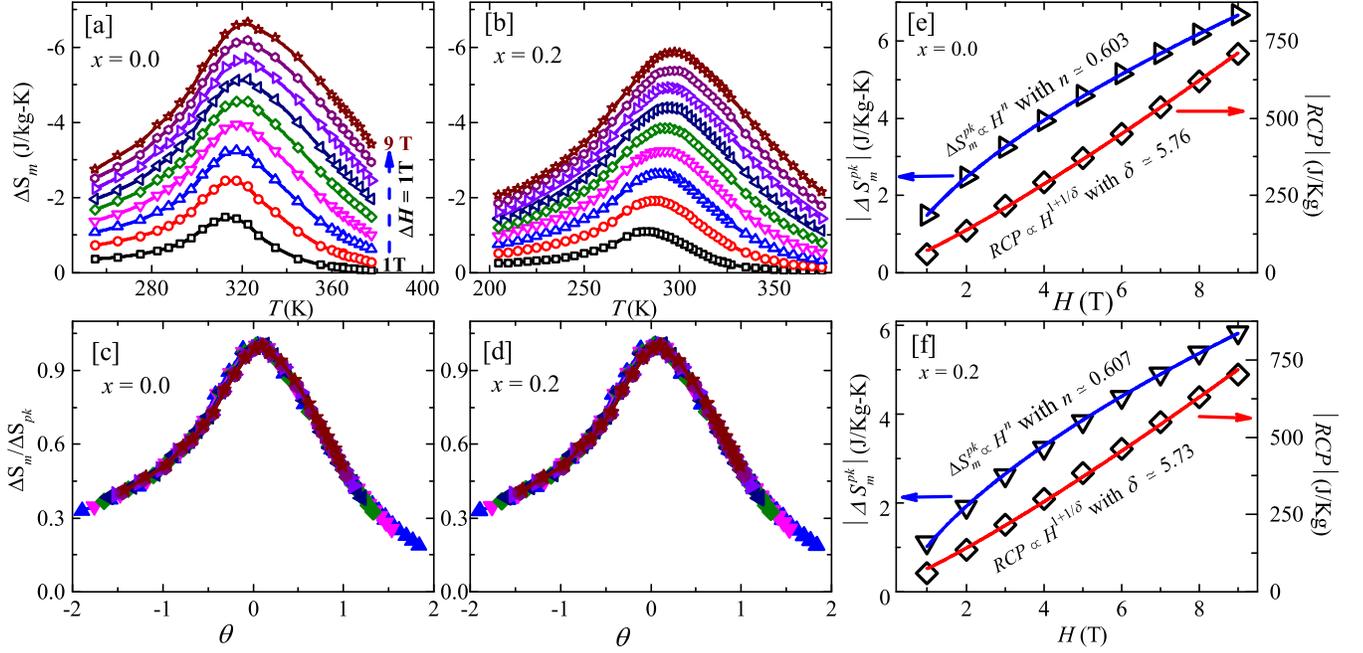}
	\caption{\label{Fig12} Temperature dependent magnetic entropy change ($\Delta S_{\rm m}$) for Mn$_{1+x}$Fe$_{4-x}$Si$_3$ [a] $x=0.0$ and [b] $x=0.2$ in the magnetic field change of 1~T to 9~T, obtained from the isothermal curves using Eq.~\eqref{eq.11}. Normalized magnetic entropy as a function of the rescaled temperature $\theta$ in different fields for [c] $x=0.0$ and [d] $x=0.2$. Magnitude of the maximum peak value of $\Delta S_{\rm m}$ ($\Delta S_{\rm m}^{pk}$) and relative cooling power ($RCP$) as a function of magnetic field in the left and right $y$-axes, respectively for [e] $x=0.0$ and [f] $x=0.2$. Both the quantities are fitted by the corresponding scaling equation.}
\end{figure}

\subsection*{Magnetocaloric Effect}
As we have seen earlier, Mn substitution tunes the value of $\Theta_{\rm CW}$ and hence the $T_{\rm C}$, continuously from 328~K to 212~K as $x$ varies from 0 to 1. This type of materials are favorable for continuous magnetic refrigeration purpose. Therefore, the magnetocaloric effect (MCE) in terms of isothermal change in magnetic entropy ($\Delta S_{\rm m}$) is studied for two compositions ($x = 0.0$ and 0.2). From the magnetization isotherms ($M$ vs $H$) at various temperatures, $\Delta S_{\rm m}$ values are calculated using the Maxwell relation: 
\begin{equation}\label{eq.11}
	\Delta S_{m}=\int_{H_{i}}^{H_{f}}\dfrac{dM}{dT}dH.
\end{equation}
Figure~\ref{Fig12}[a] and [b] present the temperature variation of $\Delta S_{\rm m}$ around the PM-FM transition at different magnetic fields up to 9~T for $x = 0.0$  and 0.2, respectively. Both the compounds show a large negative conventional MCE ($\Delta S_{\rm m}$) with a maxima at the transition temperature. This is a typical caret-like shape, akin to second order magnetic transition for both compositions.\cite{Tishin2016}
For the parent compound ($x = 0.0$), maximum value of $\Delta S_{\rm m}$ ($\sim-2$~J/kg-K) for a field change ($\Delta H$) of 2~T is found to match with the previous reports.\cite{Songlin249,Gourdon56,Hering7128,Herlitschke094304}
%From the spin dynamics studies using neutron scattering experiments, Biniskos~et~al\cite{Biniskos104407} pointed out that the critical fluctuations in the parent compound are largely responsible for enhanced MCE which are suppressed in a magnetic field of 2~T giving optimum value of MCE. In contrast, from our experiments it is found that MCE ($\Delta S_{\rm m}$) is continuously increasing with magnetic field without any signature of saturation upto 9~T. This indicates that the critical fluctuations are not the only reason for large MCE in the present compounds. The other possible ingredients for large MCE could be strong magnetocrystalline anisotropy, preferential occupancy of Mn/Fe atoms etc., which can not be assessed from the present data on the polycrystalline sample.
The $\Delta S_{\rm m}$ reaches a maximum value of $\sim-6.67$~J/Jg-K and $\sim-5.84$~J/Jg-K at their respective $T_{\rm C}$s for a field change of 9~T for $x=0$ and 0.2 compositions, respectively. Slightly smaller value of $\Delta S_{\rm m}$ for the doped samples could be due to a small reduction in magnetic moment with Mn substitution. Although these values are lower than the well known magneto-caloric material such as Gd, MnAs, Gd$_5$Si$_2$Ge$_2$, FeRh etc, but comparable with other materials showing standard MCE across the second order magnetic transition, near room temperature.\cite{GschneidnerJr1479,Franco112,Moya439} The possible reasons for enhanced MCE in these materials could be the strong magnetocrystalline anisotropy, preferential occupancy of Mn/Fe atoms etc, which can not be assessed from the present data on the polycrystalline sample

In addition, MCE is also being utilized to study the critical phenomena and the nature of the magnetic phase transition from the scaling behavior of $\Delta S_{\rm m}$.\cite{Law2680,Franco414004} The phenomenological universal scaling curve construction was first proposed by Franco~et~al\cite{Franco222512,Franco093903} in 2006 which was later utilized for analyzing the nature of magnetic phase transitions.\cite{Bonilla224424} More recently, critical analysis of MCE has also been carried out quantitatively and proven to be very effective for a detail understanding of the magnetic phase transition.\cite{Law2680} Here, we have performed the universal curve construction and the critical analysis of MCE for both $x = 0.0$ and 0.2 samples following the procedure described in Refs.~\cite{Franco222512,Franco093903}. In the universal curve construction, magnetic entropy curve is normalized to its maximum peak value [$\Delta S_{\rm m}(T)/\Delta S_{\rm m}^{\rm pk}$] at each $\Delta H$ value and is plotted as a function of rescaled temperature $\theta$. To define $\theta$, we first choose two reference temperatures ($T_{\rm r1}$ and $T_{\rm r2}$) which must satisfy the condition: $\Delta S_{\rm m}(T_{\rm r1} < T_{\rm C})/\Delta S_{\rm m}^{\rm pk} = \Delta S_{\rm m}(T_{\rm r2} > T_{\rm C})/\Delta S_{\rm m}^{\rm pk} = h$ where $h$ is a constant which has a value within the range $0 < h < 1$. The rescaled temperature can be calculated as,
\begin{equation}\label{eq.12}
	\theta=\begin{cases}
		-(T-T_{\rm C})/(T_{\rm r1}-T_{\rm C}), & \text{if $T \le T_{\rm C}$}\\
		(T-T_{\rm C})/(T_{\rm r2}-T_{\rm C}), & \text{if $T > T_{\rm C}$}.
	\end{cases}
\end{equation}
In our system, we have taken $T_{\rm C} = 309.6$~K and 278~K for $x = 0.0$ and 0.2, respectively obtained from the critical analysis of magnetization and $T_{\rm r1}$ and $T_{\rm r2}$ values are chosen corresponding to $h = 0.5$. It is reported that for materials whose $T_{\rm C}$ is near room temperature, scaling laws at the $T_{\rm C}$ are applicable for $\Delta H$ as high as $\sim 10$~T.\cite{Carlos134401} Thus, for our systems, one can apply scaling laws in the measured field range upto 9~T. 
Figure~\ref{Fig12}[c] and [d] present the $\Delta S_{\rm m}(T)/\Delta S_{\rm m}^{\rm pk}$ vs $\theta$ curves for $x = 0.0$ and 0.2, respectively for different values of $\Delta H$. It is quite apparent that all the normalized entropy curves with various $\Delta H$ values collapse into a single curve for both the compositions. This behavior is similar to the universal $\Delta S_{\rm m}$ curve reported for other compounds with second order magnetic phase transition.\cite{Bonilla224424}
\begin{figure}[htbp]
	\begin{center}
		\includegraphics[width = 7 cm,height = 9.5 cm]{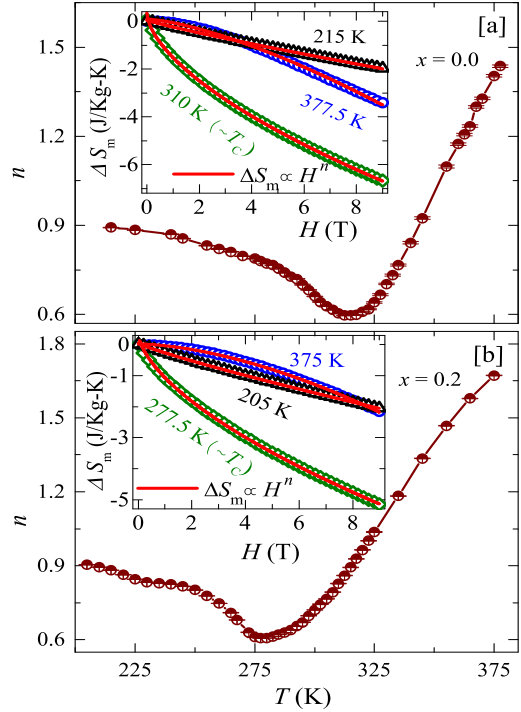}
	\end{center}
	\caption{\label{Fig13} The exponent $n$ as function of temperature obtained from the fitting of field dependent isothermal magnetic entropy change at various temperatures for [a] $x=0.0$ and [b] $x = 0.2$. Insets: field dependent isothermal magnetic entropy change $\Delta S_{\rm m}$ at three different temperatures, near $T_{\rm C}$. The solid lines are the fits using power law, as described in the text.}
\end{figure}

From the $\Delta S_{\rm m}$ vs $T$ data, the relative cooling power ($RCP$) for each $\Delta H$ value is calculated as the product of $\Delta S_{\rm m}^{\rm pk}$ and the full width at half maxima (FWHM). Figure~\ref{Fig12}[e] and [f] show the plot of $\Delta S_{\rm m}^{\rm pk}$ and $RCP$ as a function of magnetic field in the left and right $y$-axes, respectively for $x= 0.0$ and 0.2 samples. Both the quantities are found to increase with increasing magnetic field change. At the highest measured field $\Delta H = 9$~T, the $RCP$ value reaches $RCP \simeq 707$~J/kg. For the purpose of critical analysis, we have fitted these magnetic field dependent curves ($\Delta S_{\rm m}^{\rm pk}$ and $RCP$) using the following power laws\cite{Franco222512,Franco093903,Law2680}
\begin{equation}\label{eq.13}
	\lvert\Delta S_{m}^{pk}\rvert\propto H^{n},
\end{equation}
where, $n$ is a temperature dependent parameter and related to the critical exponents $\beta$ and $\gamma$ at/near the $T_{\rm C}$ as
\begin{equation}
	n=1 +\dfrac{\beta-1}{\beta+\gamma}
	\label{eq.14}
\end{equation}
and
\begin{equation}
	RCP\propto H^{1 + 1/\delta}.
	\label{eq.15}
\end{equation}
The fit of $\Delta S_{\rm m}^{\rm pk} (H)$ data by Eq.~\eqref{eq.13} yields $n \simeq 0.603$ and 0.607 for $x = 0.0$ and 0.2, respectively. They are in very good agreement with the values obtained from Eq.~\eqref{eq.14} using the $\beta$ and $\gamma$ values from the MAPs and KF plots (Table~\ref{arttype}). This proves the robustness of the critical analysis method. Similarly, the field dependent $RCP(H)$ data are fitted by Eq.~\eqref{eq.15} which gives the critical exponent value $ \delta \simeq 5.76 $ and $ 5.73 $ for $ x = 0.0 $ and $ 0.2$, respectively. These are of course very close to the $ \delta $ values obtained from critical analysis of magnetic isotherms (see Table~\ref{arttype}).

For a more quantitative analysis of MCE, we fitted the field dependent isothermal magnetic entropy change $\Delta S_{\rm m}(H)$ at various temperatures across the PM-FM transition using the power law
$\Delta S_{m}\propto H^{n}$.\cite{Franco222512}
The obtained exponent $n$ is plotted as a function of temperature in Fig.~\ref{Fig13} [a] and [b] for compositions $x = 0.0$ and 0.2, respectively. Inset of Fig.~\ref{Fig13} [a] and [b] present $\Delta S_{\rm m}$ vs $H$ plots at three different temperatures: one at low temperature ($T<T_{\rm C}$), one close to critical regime ($T\sim T_{\rm C}$), and another at high temperature ($T>T_{\rm C}$). It can be seen that for $T<T_{\rm C}$, $\Delta S_{\rm m}$ exhibits almost a linear behavior with $H$ and the exponent $n$ is found to be $\sim 0.9$ for both compositions, which is close to 1. The value $n \sim 1$ suggests that the term $\Big(\dfrac{dM}{dT}\Big)$ in Eq.~\eqref{eq.11} is weakly field dependent at low temperatures ($T<T_{\rm C}$).
Further, with rise in temperature, $n$ decreases and arrives a minimum value of 0.604 and 0.607 at $T \sim T_{\rm C}$ for compositions $x=0.0$ and 0.2, respectively. These $n$ values are consistent with the values obtained from the analysis of $\Delta S_{\rm m}^{\rm pk}$ vs $H$ using power law as shown Fig.~\ref{Fig12} [e] and [f] and also from Eq.~(\ref{eq.14}) using appropriate values of $\beta$ and $\gamma$.
Above $T_{\rm C}$, $n$ increases almost linearly and reaches a maximum value of $\sim 1.436$ and $\sim 1.671$ for compositions $x= 0.0$ and 0.2, respectively at the highest measured temperature.

The overall temperature dependence of $n$ is quite similar to that observed for other compounds showing second order magnetic phase transition.\cite{Franco222512,Franco414004} Recently, Law et al,\cite{Law2680} simulated the temperature variation of $n$ using the Bean and Rodbell model and  showed that one can quantitatively distinguish the first order and second order phase transitions by measuring $n(T)$ which was also experimentally verified by them. According to them, for a second order magnetic phase transition, $n(T)$ should exhibit a minima near $T_{\rm C}$ and for $T > T_{\rm C}$ it should increase systematically upto a maximum value of 2. Indeed, our experimental $n(T)$ behaviour for both the compositions matches well with the above predictions, confirming the second order nature of the magnetic phase transition.

\section*{Summary}
We have done a detailed investigation of the PM-FM phase transition in Mn$_{1+x}$Fe$_{4-x}$Si$_{3}$ series. A careful magnetization measurement on the parent compound rules out the presence of thermal hysteresis, establishing the second order nature of the transition. This is in contrast with the previous reports.\cite{Hering7128} This PM-FM transition is found to be tuned from $\sim 328$~K to $\sim 212$~K by Mn substitution at the Fe site upto $x=1$. We did not observe any signature of Landau tricritical point as predicted earlier for the parent compound.\cite{Herlitschke094304} Though, our temperature dependent powder XRD for $x = 1$ reveals no structural transition down to 15~K but the temperature variation of lattice parameters point towards a lattice distortion across the magnetic transition ($T_{\rm C} \simeq 212$~K), similar to the parent compound.\cite{Hering7128} This indicates that the structural degree of freedom is weakly coupled with the spin degree of freedom in this series.

A detailed critical analysis of the magnetization data across the transition is carried out for two compositions $x = 0.0$ and 0.2.
The critical exponents are estimated to be ($\beta=0.308$ and $\gamma=1.448$ from MAPs and $\delta=5.64$ from critical isotherm) and ($\beta=0.308$ and $\gamma=1.445$ from MAPs and $\delta=5.64$ from critical isotherm) for $x = 0.0$ and 0.2, respectively. These values are further confirmed from various analysis methods and Widom scaling relations indicating the robustness of critical analysis technique. The obtained critical exponents do not fall in any of the existing standard universality class and are similar to that observed for Cr$_{75}$Fe$_{25}$ and Cr$_{70}$Fe$_{30}$.\cite{Fischer064443} However, the similar values of critical exponents for both parent and doped compounds indicates that the universality class of the compound does not change and the spin-spin interaction mechanism remains unaltered upon Mn substitution. The effective critical exponents ($\beta_{\rm eff}$ and $\gamma_{\rm eff}$) seem to approach the actual experimental values in the asymptotic regime ($\epsilon \rightarrow 0$). The reliability of the critical exponents and the value of $T_{\rm C}$ are further confirmed from the scaling of magnetization, where all magnetic isotherms fall into two separate branches: one above and another below the $T_{\rm C}$. Furthermore, these critical exponents are identical to the ones obtained from the renormalization group theory calculation for $d=2$, $n=1$, and $\sigma=1.41$, which indicates long-range interactions between magnetic spins and it decays following $J(r)\sim r^{-3.41}$.

A reasonably large and negative MCE is inferred for the parent compound across the magnetic transition from the calculation of $\Delta S_{\rm m}$ vs $T$. Upon Mn substitution at the Fe site, the magnitude of $\Delta S_{\rm m}$ at the peak position is reduced slightly which is likely due to the reduction in magnetic moment. On the other hand, the value of $\Delta S_{\rm m}$ at the peak position is found to be enhanced continuously with magnetic field, for both the compounds. The maximum estimated value of $\Delta S_{\rm m}$ is found to be -6.67~J/Kg-K and -5.84~J/Kg-K in a field change of 9~T for $x = 0.0$ and 0.2, respectively. Interestingly, a large and same value of $RCP$ ($\sim 707$~J/Kg) was found for both the compositions in a field change of 9~T. The universal scaling of MCE shows that the $\Delta S_{\rm m}(T)$ curves for different $\Delta H$ values collapse on the master curve for both the compositions. The obtained critical exponents ($n$ and $\delta$) from the critical analysis of field dependent $\Delta S_{\rm m}$ and $RCP$ are in good agreement with the other analysis results. The second order character of PM-FM transition, MCE, and $RCP$ in the parent compound are also consistent with the recent Monte-Carlo studies.\cite{Bouachraoui151785} Thus, the tunability of the PM-FM transition with Mn substitution and its reversible character make MnFe$_4$Si$_3$ a potential candidate for magnetic refrigeration application.

%\bibliography{ref_MnFe4Si3}

\begin{thebibliography}{10}
	\urlstyle{rm}
	\expandafter\ifx\csname url\endcsname\relax
	\def\url#1{\texttt{#1}}\fi
	\expandafter\ifx\csname urlprefix\endcsname\relax\def\urlprefix{URL }\fi
	\expandafter\ifx\csname doiprefix\endcsname\relax\def\doiprefix{DOI: }\fi
	\providecommand{\bibinfo}[2]{#2}
	\providecommand{\eprint}[2][]{\url{#2}}
	
	\bibitem{Franco112}
	\bibinfo{author}{Franco, V.} \emph{et~al.}
	\newblock \bibinfo{journal}{\bibinfo{title}{Magnetocaloric effect: From
			materials research to refrigeration devices}}.
	\newblock {\emph{\JournalTitle{Prog. Mater. Sci.}}}
	\textbf{\bibinfo{volume}{93}}, \bibinfo{pages}{112},
	\doiprefix\url{10.1016/j.pmatsci.2017.10.005} (\bibinfo{year}{2018}).
	
	\bibitem{GschneidnerJr1479}
	\bibinfo{author}{Gschneidner~Jr, K.~A.}, \bibinfo{author}{Pecharsky, V.~K.} \&
	\bibinfo{author}{Tsokol, A.~O.}
	\newblock \bibinfo{journal}{\bibinfo{title}{Recent developments in
			magnetocaloric materials}}.
	\newblock {\emph{\JournalTitle{Rep. Prog. Phys.}}}
	\textbf{\bibinfo{volume}{68}}, \bibinfo{pages}{1479},
	\doiprefix\url{10.1088/0034-4885/68/6/r04} (\bibinfo{year}{2005}).
	
	\bibitem{Tishin2016}
	\bibinfo{author}{Tishin, A.} \& \bibinfo{author}{Spichkin, Y.}
	\newblock \emph{\bibinfo{title}{The Magnetocaloric Effect and its
			Applications}} (\bibinfo{publisher}{{CRC} Press}, \bibinfo{year}{2016}).
	
	\bibitem{Pecharsky44}
	\bibinfo{author}{Pecharsky, V.~K.} \& \bibinfo{author}{Gschneidner~Jr, K.~A.}
	\newblock \bibinfo{journal}{\bibinfo{title}{Magnetocaloric effect and magnetic
			refrigeration}}.
	\newblock {\emph{\JournalTitle{J. Magn. Magn. Mater.}}}
	\textbf{\bibinfo{volume}{200}}, \bibinfo{pages}{44},
	\doiprefix\url{10.1016/S0304-8853(99)00397-2} (\bibinfo{year}{1999}).
	
	\bibitem{Pecharsky4494}
	\bibinfo{author}{Pecharsky, V.~K.} \& \bibinfo{author}{Gschneidner, K.~A., Jr.}
	\newblock \bibinfo{journal}{\bibinfo{title}{Giant magnetocaloric effect in
			{Gd$_5$Si$_2$Ge$_2$}}}.
	\newblock {\emph{\JournalTitle{Phys. Rev. Lett.}}}
	\textbf{\bibinfo{volume}{78}}, \bibinfo{pages}{4494},
	\doiprefix\url{10.1103/PhysRevLett.78.4494} (\bibinfo{year}{1997}).
	
	\bibitem{Gschneidner387}
	\bibinfo{author}{Gschneidner, K.~A.} \& \bibinfo{author}{Pecharsky, V.~K.}
	\newblock \bibinfo{journal}{\bibinfo{title}{Magnetocaloric materials}}.
	\newblock {\emph{\JournalTitle{Annu. Rev. Mater. Sci.}}}
	\textbf{\bibinfo{volume}{30}}, \bibinfo{pages}{387},
	\doiprefix\url{10.1146/annurev.matsci.30.1.387} (\bibinfo{year}{2000}).
	
	\bibitem{Moya439}
	\bibinfo{author}{Moya, X.}, \bibinfo{author}{Kar-Narayan, S.} \&
	\bibinfo{author}{Mathur, N.~D.}
	\newblock \bibinfo{journal}{\bibinfo{title}{Caloric materials near ferroic
			phase transitions}}.
	\newblock {\emph{\JournalTitle{Nat. Mater.}}} \textbf{\bibinfo{volume}{13}},
	\bibinfo{pages}{439}, \doiprefix\url{10.1038/nmat3951}
	(\bibinfo{year}{2014}).
	
	\bibitem{Roy183201}
	\bibinfo{author}{Roy, S.~B.}
	\newblock \bibinfo{journal}{\bibinfo{title}{First order magneto-structural
			phase transition and associated multi-functional properties in magnetic
			solids}}.
	\newblock {\emph{\JournalTitle{J. Phys.: Condens Mater}}}
	\textbf{\bibinfo{volume}{25}}, \bibinfo{pages}{183201},
	\doiprefix\url{10.1088/0953-8984/25/18/183201} (\bibinfo{year}{2013}).
	
	\bibitem{Tegus150}
	\bibinfo{author}{Tegus, O.}, \bibinfo{author}{Br{\"u}ck, E.},
	\bibinfo{author}{Buschow, K. H.~J.} \& \bibinfo{author}{de~Boer, F.~R.}
	\newblock \bibinfo{journal}{\bibinfo{title}{Transition-metal-based magnetic
			refrigerants for room-temperature applications}}.
	\newblock {\emph{\JournalTitle{Nature}}} \textbf{\bibinfo{volume}{415}},
	\bibinfo{pages}{150}, \doiprefix\url{10.1038/415150a} (\bibinfo{year}{2002}).
	
	\bibitem{Chaikin1995}
	\bibinfo{author}{Chaikin, P.~M.} \& \bibinfo{author}{Lubensky, T.~C.}
	\newblock \emph{\bibinfo{title}{Principles of Condensed Matter Physics}}
	(\bibinfo{publisher}{Cambridge University Press},
	\bibinfo{address}{Cambridge}, \bibinfo{year}{1995}).
	
	\bibitem{Stanley336}
	\bibinfo{author}{{Stanley}, H.~E.}
	\newblock \emph{\bibinfo{title}{{Introduction to Phase Transitions and Critical
				Phenomena}}} (\bibinfo{publisher}{Oxford University Press},
	\bibinfo{year}{1987}).
	
	\bibitem{Li152403}
	\bibinfo{author}{Li, L.} \emph{et~al.}
	\newblock \bibinfo{journal}{\bibinfo{title}{Giant reversible magnetocaloric
			effect in {ErMn$_2$Si$_2$} compound with a second order magnetic phase
			transition}}.
	\newblock {\emph{\JournalTitle{Appl.Phys.Lett.}}}
	\textbf{\bibinfo{volume}{100}}, \bibinfo{pages}{152403},
	\doiprefix\url{10.1063/1.4704155} (\bibinfo{year}{2012}).
	
	\bibitem{Singh3321}
	\bibinfo{author}{Singh, S.} \emph{et~al.}
	\newblock \bibinfo{journal}{\bibinfo{title}{Large magnetization and reversible
			magnetocaloric effect at the second-order magnetic transition in heusler
			materials}}.
	\newblock {\emph{\JournalTitle{Adv. Mater.}}} \textbf{\bibinfo{volume}{28}},
	\bibinfo{pages}{3321}, \doiprefix\url{10.1002/adma.201505571}
	(\bibinfo{year}{2016}).
	
	\bibitem{Gschneidner945}
	\bibinfo{author}{Gschneidner, K.~A.} \& \bibinfo{author}{Pecharsky, V.~K.}
	\newblock \bibinfo{journal}{\bibinfo{title}{Thirty years of near room
			temperature magnetic cooling{:} where we are today and future prospects}}.
	\newblock {\emph{\JournalTitle{Int. J. Refrig.}}}
	\textbf{\bibinfo{volume}{31}}, \bibinfo{pages}{945},
	\doiprefix\url{10.1016/j.ijrefrig.2008.01.004} (\bibinfo{year}{2008}).
	
	\bibitem{Bruck763}
	\bibinfo{author}{Br{\"u}ck, E.}, \bibinfo{author}{Tegus, O.},
	\bibinfo{author}{Cam~Thanh, D.~T.}, \bibinfo{author}{Trung, N.~T.} \&
	\bibinfo{author}{Buschow, K. H.~J.}
	\newblock \bibinfo{journal}{\bibinfo{title}{A review on mn based materials for
			magnetic refrigeration{:} structure and properties}}.
	\newblock {\emph{\JournalTitle{Int. J. Refrig.}}}
	\textbf{\bibinfo{volume}{31}}, \bibinfo{pages}{763},
	\doiprefix\url{10.1016/j.ijrefrig.2007.11.013} (\bibinfo{year}{2008}).
	
	\bibitem{Gutfleisch821}
	\bibinfo{author}{Gutfleisch, O.} \emph{et~al.}
	\newblock \bibinfo{journal}{\bibinfo{title}{Magnetic materials and devices for
			the 21st century{:} stronger, lighter, and more energy efficient}}.
	\newblock {\emph{\JournalTitle{Adv. Mater.}}} \textbf{\bibinfo{volume}{23}},
	\bibinfo{pages}{821}, \doiprefix\url{10.1002/adma.201002180}
	(\bibinfo{year}{2011}).
	
	\bibitem{Chirkova15}
	\bibinfo{author}{Chirkova, A.} \emph{et~al.}
	\newblock \bibinfo{journal}{\bibinfo{title}{Giant adiabatic temperature change
			in ferh alloys evidenced by direct measurements under cyclic conditions}}.
	\newblock {\emph{\JournalTitle{Acta Mater.}}} \textbf{\bibinfo{volume}{106}},
	\bibinfo{pages}{15}, \doiprefix\url{10.1016/j.actamat.2015.11.054}
	(\bibinfo{year}{2016}).
	
	\bibitem{Krenke450}
	\bibinfo{author}{Krenke, T.} \emph{et~al.}
	\newblock \bibinfo{journal}{\bibinfo{title}{Inverse magnetocaloric effect in
			ferromagnetic {Ni-Mn-Sn} alloys}}.
	\newblock {\emph{\JournalTitle{Nat. Mater.}}} \textbf{\bibinfo{volume}{4}},
	\bibinfo{pages}{450}, \doiprefix\url{10.1038/nmat1395}
	(\bibinfo{year}{2005}).
	
	\bibitem{Gottschilch15275}
	\bibinfo{author}{Gottschilch, M.} \emph{et~al.}
	\newblock \bibinfo{journal}{\bibinfo{title}{Study of the antiferromagnetism of
			{Mn$_5$Si$_3$}{:} an inverse magnetocaloric effect material}}.
	\newblock {\emph{\JournalTitle{J. Mater. Chem.}}}
	\textbf{\bibinfo{volume}{22}}, \bibinfo{pages}{15275},
	\doiprefix\url{10.1039/C2JM00154C} (\bibinfo{year}{2012}).
	
	\bibitem{Gourdon56}
	\bibinfo{author}{Gourdon, O.} \emph{et~al.}
	\newblock \bibinfo{journal}{\bibinfo{title}{Toward a better understanding of
			the magnetocaloric effect: An experimental and theoretical study of
			{MnFe$_4$Si$_3$}}}.
	\newblock {\emph{\JournalTitle{J. Solid State Chem.}}}
	\textbf{\bibinfo{volume}{216}}, \bibinfo{pages}{56},
	\doiprefix\url{10.1016/j.jssc.2014.05.001} (\bibinfo{year}{2014}).
	
	\bibitem{Songlin249}
	\bibinfo{author}{Songlin, D.} \emph{et~al.}
	\newblock \bibinfo{journal}{\bibinfo{title}{Magnetic phase transition and
			magnetocaloric effect in {Mn$_{5-x}$Fe$_x$Si$_3$}}}.
	\newblock {\emph{\JournalTitle{J. Alloys Compd.}}}
	\textbf{\bibinfo{volume}{334}}, \bibinfo{pages}{249},
	\doiprefix\url{10.1016/S0925-8388(01)01776-5} (\bibinfo{year}{2002}).
	
	\bibitem{Johnson311}
	\bibinfo{author}{Johnson, V.}, \bibinfo{author}{Weiher, J.~F.},
	\bibinfo{author}{Frederick, C.~G.} \& \bibinfo{author}{Rogers, D.~B.}
	\newblock \bibinfo{journal}{\bibinfo{title}{Magnetic and m{\"o}ssbauer effect
			studies of {Mn$_5$Si$_3$:Fe$_5$Si$_3$} solid solutions}}.
	\newblock {\emph{\JournalTitle{J. Solid State Chem.}}}
	\textbf{\bibinfo{volume}{4}}, \bibinfo{pages}{311},
	\doiprefix\url{10.1016/0022-4596(72)90122-3} (\bibinfo{year}{1972}).
	
	\bibitem{Candini6819}
	\bibinfo{author}{Candini, A.} \emph{et~al.}
	\newblock \bibinfo{journal}{\bibinfo{title}{Revised magnetic phase diagram for
			{Fe$_x$Mn$_{5-x}$Si$_3$} intermetallics}}.
	\newblock {\emph{\JournalTitle{J. Appl. Phys.}}} \textbf{\bibinfo{volume}{95}},
	\bibinfo{pages}{6819}, \doiprefix\url{10.1063/1.1688219}
	(\bibinfo{year}{2004}).
	
	\bibitem{Brown7619}
	\bibinfo{author}{Brown, P.~J.} \& \bibinfo{author}{Forsyth, J.~B.}
	\newblock \bibinfo{journal}{\bibinfo{title}{Antiferromagnetism in
			{Mn$_5$Si$_3$}: the magnetic structure of the {AF2} phase at {$70$~K}}}.
	\newblock {\emph{\JournalTitle{J. Phys.: Condens. Matter}}}
	\textbf{\bibinfo{volume}{7}}, \bibinfo{pages}{7619},
	\doiprefix\url{10.1088/0953-8984/7/39/004} (\bibinfo{year}{1995}).
	
	\bibitem{Surgers055604}
	\bibinfo{author}{S{\"u}rgers, C.}, \bibinfo{author}{Kittler, W.},
	\bibinfo{author}{Wolf, T.} \& \bibinfo{author}{L{\"o}hneysen, H.~v.}
	\newblock \bibinfo{journal}{\bibinfo{title}{Anomalous hall effect in the
			noncollinear antiferromagnet {Mn$_5$Si$_3$}}}.
	\newblock {\emph{\JournalTitle{AIP Advances}}} \textbf{\bibinfo{volume}{6}},
	\bibinfo{pages}{055604}, \doiprefix\url{10.1063/1.4943759}
	(\bibinfo{year}{2016}).
	
	\bibitem{Surgers3400}
	\bibinfo{author}{S{\"u}rgers, C.}, \bibinfo{author}{Fischer, G.},
	\bibinfo{author}{Winkel, P.} \& \bibinfo{author}{L{\"o}hneysen, H.~v.}
	\newblock \bibinfo{journal}{\bibinfo{title}{Large topological hall effect in
			the non-collinear phase of an antiferromagnet}}.
	\newblock {\emph{\JournalTitle{Nat. Commun.}}} \textbf{\bibinfo{volume}{5}},
	\bibinfo{pages}{3400}, \doiprefix\url{10.1038/ncomms4400}
	(\bibinfo{year}{2014}).
	\newblock \bibinfo{note}{Article}.
	
	\bibitem{Biniskos257205}
	\bibinfo{author}{Biniskos, N.} \emph{et~al.}
	\newblock \bibinfo{journal}{\bibinfo{title}{Spin fluctuations drive the inverse
			magnetocaloric effect in {Mn$_5$Si$_3$}}}.
	\newblock {\emph{\JournalTitle{Phys. Rev. Lett.}}}
	\textbf{\bibinfo{volume}{120}}, \bibinfo{pages}{257205},
	\doiprefix\url{10.1103/PhysRevLett.120.257205} (\bibinfo{year}{2018}).
	
	\bibitem{Shinjo797}
	\bibinfo{author}{Shinjo, T.}, \bibinfo{author}{Nakamura, Y.} \&
	\bibinfo{author}{Shikazono, N.}
	\newblock \bibinfo{journal}{\bibinfo{title}{Magnetic study of {Fe$_3$Si} and
			{Fe$_5$Si$_3$} by m{\"o}ssbauer effect}}.
	\newblock {\emph{\JournalTitle{J. Phys. Soc. Jpn.}}}
	\textbf{\bibinfo{volume}{18}}, \bibinfo{pages}{797},
	\doiprefix\url{10.1143/JPSJ.18.797} (\bibinfo{year}{1963}).
	
	\bibitem{Johnson465}
	\bibinfo{author}{Johnson, C.~E.}, \bibinfo{author}{Forsyth, J.~B.},
	\bibinfo{author}{Lander, G.~H.} \& \bibinfo{author}{Brown, P.~J.}
	\newblock \bibinfo{journal}{\bibinfo{title}{Magnetic moments and hyperfine
			interactions in carbon-stabilized {Fe$_5$Si$_3$}}}.
	\newblock {\emph{\JournalTitle{J. Appl. Phys.}}} \textbf{\bibinfo{volume}{39}},
	\bibinfo{pages}{465}, \doiprefix\url{10.1063/1.2163482}
	(\bibinfo{year}{1968}).
	
	\bibitem{Narasimhan1511}
	\bibinfo{author}{Narasimhan, K. S. V.~L.}, \bibinfo{author}{Reiff, W.~M.},
	\bibinfo{author}{Steinfink, H.} \& \bibinfo{author}{Collins, R.~L.}
	\newblock \bibinfo{journal}{\bibinfo{title}{Magnetism and bonding in a {D88}
			structure; m{\"o}ssbauer and magnetic investigation of the system
			{Mn$_5$Si$_3$-Fe$_5$Si$_3$}}}.
	\newblock {\emph{\JournalTitle{J. Phys. Chem. Solids}}}
	\textbf{\bibinfo{volume}{31}}, \bibinfo{pages}{1511},
	\doiprefix\url{10.1016/0022-3697(70)90035-1} (\bibinfo{year}{1970}).
	
	\bibitem{Binczycka_K13}
	\bibinfo{author}{Binczycka, H.}, \bibinfo{author}{Dimitrijevic, Z.},
	\bibinfo{author}{Gajic, B.} \& \bibinfo{author}{Szytula, A.}
	\newblock \bibinfo{journal}{\bibinfo{title}{Atomic and magnetic structure of
			{Mn$_{5-x}$Fe$_x$Si$_3$}}}.
	\newblock {\emph{\JournalTitle{Phys. Stat. Solidi A}}}
	\textbf{\bibinfo{volume}{19}}, \bibinfo{pages}{K13},
	\doiprefix\url{10.1002/pssa.2210190145} (\bibinfo{year}{1973}).
	
	\bibitem{Hering7128}
	\bibinfo{author}{Hering, P.} \emph{et~al.}
	\newblock \bibinfo{journal}{\bibinfo{title}{Structure, magnetism, and the
			magnetocaloric effect of {MnFe$_4$Si$_3$} single crystals and powder
			samples}}.
	\newblock {\emph{\JournalTitle{Chem. Mater.}}} \textbf{\bibinfo{volume}{27}},
	\bibinfo{pages}{7128}, \doiprefix\url{10.1021/acs.chemmater.5b03123}
	(\bibinfo{year}{2015}).
	
	\bibitem{Haug539}
	\bibinfo{author}{Haug, R.}, \bibinfo{author}{Kappel, G.} \&
	\bibinfo{author}{Jaegle, A.}
	\newblock \bibinfo{journal}{\bibinfo{title}{Electrical resistivity and magnetic
			susceptibility studies of the system {Mn$_5$Si$_3$-Fe$_5$Si$_3$}}}.
	\newblock {\emph{\JournalTitle{J. Phys. Chem. Solids}}}
	\textbf{\bibinfo{volume}{41}}, \bibinfo{pages}{539},
	\doiprefix\url{10.1016/0022-3697(80)90003-7} (\bibinfo{year}{1980}).
	
	\bibitem{Herlitschke094304}
	\bibinfo{author}{Herlitschke, M.} \emph{et~al.}
	\newblock \bibinfo{journal}{\bibinfo{title}{Elasticity and magnetocaloric
			effect in {MnFe$_4$Si$_3$}}}.
	\newblock {\emph{\JournalTitle{Phys. Rev. B}}} \textbf{\bibinfo{volume}{93}},
	\bibinfo{pages}{094304}, \doiprefix\url{10.1103/PhysRevB.93.094304}
	(\bibinfo{year}{2016}).
	
	\bibitem{Biniskos104407}
	\bibinfo{author}{Biniskos, N.} \emph{et~al.}
	\newblock \bibinfo{journal}{\bibinfo{title}{Spin dynamics of the magnetocaloric
			compound {MnFe$_4$Si$_3$}}}.
	\newblock {\emph{\JournalTitle{Phys. Rev. B}}} \textbf{\bibinfo{volume}{96}},
	\bibinfo{pages}{104407}, \doiprefix\url{10.1103/PhysRevB.96.104407}
	(\bibinfo{year}{2017}).
	
	\bibitem{Sinha072017}
	\bibinfo{author}{Sinha, A.~K.} \emph{et~al.}
	\newblock \bibinfo{journal}{\bibinfo{title}{Angle dispersive x-ray diffraction
			beamline on indus-2 synchrotron radiation source: Commissioning and first
			results}}.
	\newblock {\emph{\JournalTitle{J. Phys. Conf. Ser.}}}
	\textbf{\bibinfo{volume}{425}}, \bibinfo{pages}{072017},
	\doiprefix\url{10.1088/1742-6596/425/7/072017} (\bibinfo{year}{2013}).
	
	\bibitem{Rodriguez55}
	\bibinfo{author}{Rodríguez-Carvajal, J.}
	\newblock \bibinfo{journal}{\bibinfo{title}{Recent advances in magnetic
			structure determination by neutron powder diffraction}}.
	\newblock {\emph{\JournalTitle{Physica B}}} \textbf{\bibinfo{volume}{192}},
	\bibinfo{pages}{55}, \doiprefix\url{10.1016/0921-4526(93)90108-I}
	(\bibinfo{year}{1993}).
	
	\bibitem{Kaul1114}
	\bibinfo{author}{Kaul, S.~N.} \& \bibinfo{author}{Srinath, S.}
	\newblock \bibinfo{journal}{\bibinfo{title}{Gadolinium: A helical
			antiferromagnet or a collinear ferromagnet}}.
	\newblock {\emph{\JournalTitle{Phys. Rev. B}}} \textbf{\bibinfo{volume}{62}},
	\bibinfo{pages}{1114}, \doiprefix\url{10.1103/PhysRevB.62.1114}
	(\bibinfo{year}{2000}).
	
	\bibitem{Vegard3161}
	\bibinfo{author}{Denton, A.~R.} \& \bibinfo{author}{Ashcroft, N.~W.}
	\newblock \bibinfo{journal}{\bibinfo{title}{Vegard's law}}.
	\newblock {\emph{\JournalTitle{Phys. Rev. A}}} \textbf{\bibinfo{volume}{43}},
	\bibinfo{pages}{3161}, \doiprefix\url{10.1103/PhysRevA.43.3161}
	(\bibinfo{year}{1991}).
	
	\bibitem{Fischer064443}
	\bibinfo{author}{Fischer, S.~F.}, \bibinfo{author}{Kaul, S.~N.} \&
	\bibinfo{author}{Kronm{\"u}ller, H.}
	\newblock \bibinfo{journal}{\bibinfo{title}{Critical magnetic properties of
			disordered polycrystalline {Cr$_{75}$Fe$_{25}$} and {Cr$_{70}$Fe$_{30}$}
			alloys}}.
	\newblock {\emph{\JournalTitle{Phys. Rev. B}}} \textbf{\bibinfo{volume}{65}},
	\bibinfo{pages}{064443}, \doiprefix\url{10.1103/PhysRevB.65.064443}
	(\bibinfo{year}{2002}).
	
	\bibitem{Pramanik214426}
	\bibinfo{author}{Pramanik, A.~K.} \& \bibinfo{author}{Banerjee, A.}
	\newblock \bibinfo{journal}{\bibinfo{title}{Critical behavior at paramagnetic
			to ferromagnetic phase transition in {Pr$_{0.5}$Sr$_{0.5}$MnO$_3$}: A bulk
			magnetization study}}.
	\newblock {\emph{\JournalTitle{Phys. Rev. B}}} \textbf{\bibinfo{volume}{79}},
	\bibinfo{pages}{214426}, \doiprefix\url{10.1103/PhysRevB.79.214426}
	(\bibinfo{year}{2009}).
	
	\bibitem{Arrott1394}
	\bibinfo{author}{Arrott, A.}
	\newblock \bibinfo{journal}{\bibinfo{title}{Criterion for ferromagnetism from
			observations of magnetic isotherms}}.
	\newblock {\emph{\JournalTitle{Phys. Rev.}}} \textbf{\bibinfo{volume}{108}},
	\bibinfo{pages}{1394}, \doiprefix\url{10.1103/PhysRev.108.1394}
	(\bibinfo{year}{1957}).
	
	\bibitem{Arrott786}
	\bibinfo{author}{Arrott, A.} \& \bibinfo{author}{Noakes, J.~E.}
	\newblock \bibinfo{journal}{\bibinfo{title}{Approximate equation of state for
			nickel near its critical temperature}}.
	\newblock {\emph{\JournalTitle{Phys. Rev. Lett.}}}
	\textbf{\bibinfo{volume}{19}}, \bibinfo{pages}{786},
	\doiprefix\url{10.1103/PhysRevLett.19.786} (\bibinfo{year}{1967}).
	
	\bibitem{Banerjee16}
	\bibinfo{author}{Banerjee, B.~K.}
	\newblock \bibinfo{journal}{\bibinfo{title}{On a generalised approach to first
			and second order magnetic transitions}}.
	\newblock {\emph{\JournalTitle{Phys. Lett.}}} \textbf{\bibinfo{volume}{12}},
	\bibinfo{pages}{16}, \doiprefix\url{10.1016/0031-9163(64)91158-8}
	(\bibinfo{year}{1964}).
	
	\bibitem{KouvelA1626}
	\bibinfo{author}{Kouvel, J.~S.} \& \bibinfo{author}{Fisher, M.~E.}
	\newblock \bibinfo{journal}{\bibinfo{title}{Detailed magnetic behavior of
			nickel near its curie point}}.
	\newblock {\emph{\JournalTitle{Phys.Rev.}}} \textbf{\bibinfo{volume}{136}},
	\bibinfo{pages}{A1626}, \doiprefix\url{10.1103/PhysRev.136.A1626}
	(\bibinfo{year}{1964}).
	
	\bibitem{Widom1633}
	\bibinfo{author}{Widom, B.}
	\newblock \bibinfo{journal}{\bibinfo{title}{Degree of the critical isotherm}}.
	\newblock {\emph{\JournalTitle{J. Chem. Phys.}}} \textbf{\bibinfo{volume}{41}},
	\bibinfo{pages}{1633}, \doiprefix\url{10.1063/1.1726135}
	(\bibinfo{year}{1964}).
	
	\bibitem{Widom3898}
	\bibinfo{author}{Widom, B.}
	\newblock \bibinfo{journal}{\bibinfo{title}{Equation of state in the
			neighborhood of the critical point}}.
	\newblock {\emph{\JournalTitle{J. Chem. Phys.}}} \textbf{\bibinfo{volume}{43}},
	\bibinfo{pages}{3898}, \doiprefix\url{10.1063/1.1696618}
	(\bibinfo{year}{1965}).
	
	\bibitem{Fisher917}
	\bibinfo{author}{Fisher, M.~E.}, \bibinfo{author}{Ma, S.~K.} \&
	\bibinfo{author}{Nickel, B.~G.}
	\newblock \bibinfo{journal}{\bibinfo{title}{Critical exponents for long-range
			interactions}}.
	\newblock {\emph{\JournalTitle{Phys. Rev. Lett.}}}
	\textbf{\bibinfo{volume}{29}}, \bibinfo{pages}{917},
	\doiprefix\url{10.1103/PhysRevLett.29.917} (\bibinfo{year}{1972}).
	
	\bibitem{Kaul5}
	\bibinfo{author}{Kaul, S.~N.}
	\newblock \bibinfo{journal}{\bibinfo{title}{Static critical phenomena in
			ferromagnets with quenched disorder}}.
	\newblock {\emph{\JournalTitle{J. Magn. Magn. Mater.}}}
	\textbf{\bibinfo{volume}{53}}, \bibinfo{pages}{5},
	\doiprefix\url{10.1016/0304-8853(85)90128-3} (\bibinfo{year}{1985}).
	
	\bibitem{Law2680}
	\bibinfo{author}{Law, J.~Y.} \emph{et~al.}
	\newblock \bibinfo{journal}{\bibinfo{title}{A quantitative criterion for
			determining the order of magnetic phase transitions using the magnetocaloric
			effect}}.
	\newblock {\emph{\JournalTitle{Nat. Commun.}}} \textbf{\bibinfo{volume}{9}},
	\bibinfo{pages}{2680}, \doiprefix\url{10.1038/s41467-018-05111-w}
	(\bibinfo{year}{2018}).
	
	\bibitem{Franco414004}
	\bibinfo{author}{Franco, V.} \emph{et~al.}
	\newblock \bibinfo{journal}{\bibinfo{title}{Predicting the tricritical point
			composition of a series of {LaFeSi} magnetocaloric alloys via universal
			scaling}}.
	\newblock {\emph{\JournalTitle{J. Phys. D: Appl. Phys.}}}
	\textbf{\bibinfo{volume}{50}}, \bibinfo{pages}{414004},
	\doiprefix\url{10.1088/1361-6463/aa8792} (\bibinfo{year}{2017}).
	
	\bibitem{Franco222512}
	\bibinfo{author}{Franco, V.}, \bibinfo{author}{Bl\'{a}zquez, J.~S.} \&
	\bibinfo{author}{Conde, A.}
	\newblock \bibinfo{journal}{\bibinfo{title}{Field dependence of the
			magnetocaloric effect in materials with a second order phase transition a
			master curve for the magnetic entropy change}}.
	\newblock {\emph{\JournalTitle{Appl. Phys. Lett.}}}
	\textbf{\bibinfo{volume}{89}}, \bibinfo{pages}{222512},
	\doiprefix\url{10.1063/1.2399361} (\bibinfo{year}{2006}).
	
	\bibitem{Franco093903}
	\bibinfo{author}{Franco, V.} \emph{et~al.}
	\newblock \bibinfo{journal}{\bibinfo{title}{A constant magnetocaloric response
			in {FeMoCuB} amorphous alloys with different {Fe/B} ratios}}.
	\newblock {\emph{\JournalTitle{J. Appl. Phys.}}}
	\textbf{\bibinfo{volume}{101}}, \bibinfo{pages}{093903},
	\doiprefix\url{10.1063/1.2724804} (\bibinfo{year}{2007}).
	
	\bibitem{Bonilla224424}
	\bibinfo{author}{Bonilla, C.~M.} \emph{et~al.}
	\newblock \bibinfo{journal}{\bibinfo{title}{Universal behavior for magnetic
			entropy change in magnetocaloric materials: An analysis on the nature of
			phase transitions}}.
	\newblock {\emph{\JournalTitle{Phys. Rev. B}}} \textbf{\bibinfo{volume}{81}},
	\bibinfo{pages}{224424}, \doiprefix\url{10.1103/PhysRevB.81.224424}
	(\bibinfo{year}{2010}).
	
	\bibitem{Carlos134401}
	\bibinfo{author}{Romero-Mu\~niz, C.}, \bibinfo{author}{Tamura, R.},
	\bibinfo{author}{Tanaka, S.} \& \bibinfo{author}{Franco, V.}
	\newblock \bibinfo{journal}{\bibinfo{title}{Applicability of scaling behavior
			and power laws in the analysis of the magnetocaloric effect in second-order
			phase transition materials}}.
	\newblock {\emph{\JournalTitle{Phys. Rev. B}}} \textbf{\bibinfo{volume}{94}},
	\bibinfo{pages}{134401}, \doiprefix\url{10.1103/PhysRevB.94.134401}
	(\bibinfo{year}{2016}).
	
	\bibitem{Bouachraoui151785}
	\bibinfo{author}{Bouachraoui, R.} \emph{et~al.}
	\newblock \bibinfo{journal}{\bibinfo{title}{The magnetocaloric and magnetic
			properties of the {MnFe$_4$Si$_3$}: Monte carlo investigation}}.
	\newblock {\emph{\JournalTitle{J. Alloy. Compd.}}}
	\textbf{\bibinfo{volume}{809}}, \bibinfo{pages}{151785},
	\doiprefix\url{10.1016/j.jallcom.2019.151785} (\bibinfo{year}{2019}).
	
\end{thebibliography}
\providecommand{\noopsort}[1]{}\providecommand{\singleletter}[1]{#1}%

\section*{Acknowledgements}
 We thank Kranti K Shrama and Alok Banerjee, UGC-DAE Consortium for Scientific Research, Indore for magnetization measurements using SQUID magnetometer. We are also thankful to Archna  Sagdeo, Indus Synchrotrons Utilization Division, RRCAT, Indore for synchrotron XRD measurement. Financial support from BRNS project bearing sanction No.37(3)/14/26/2017-BRNS is acknowledged.

\section*{Author contributions statement}
V.S. synthesized the compounds, did the experiments, analyzed the data, and wrote the manuscript. P.B. performed some magnetic measurements. R.R. discussed the results and commented on the manuscript. R.N. supervised the experiments, analysis of results, and corrected the manuscript. All authors reviewed the manuscript. 

\section*{Additional information}

%To include, in this order: \textbf{Accession codes} (where applicable);
\textbf{Competing interests}: The authors declare that they have no competing interests.

\end{document}